\documentclass[final,copyright,creativecommons]{eptcs}
\providecommand{\event}{MSFP 2012}
\usepackage{amsmath}
\usepackage{xspace}
\usepackage{accents}
\usepackage[final]{listings}
\usepackage{url} % to make URLs break on line break

\newcommand*{\fatdot}[1]{\accentset{\mbox{\large\bfseries .\;}}{#1}}

\newcommand\fplus{\mathbin{:\!\!+\!\!:}}

\newcommand\fsub{\mathbin{:\!\prec\!:}}

\newcommand\fann{\mathbin{:\!\!\&\!\!:}}
\newcommand\nattrans{\mathbin{\fatdot{\rightarrow}}}

\newcommand{\dalc}{\emph{data types \`{a} la carte}\xspace}

\lstdefinelanguage{lithaskell}[]{haskell}{
 basicstyle=\scriptsize\ttfamily,
 flexiblecolumns=false,
 basewidth={0.5em,0.45em},
 literate={+}{{$+$}}1 {/}{{$/$}}1 {*}{{$*$}}1 {=}{{$=$}}1
          {>}{{$>$}}1 {<}{{$<$}}1 {\\}{{$\lambda$}}1
          {\\\\}{{\char`\\\char`\\}}1
          {->}{{$\rightarrow$}}2 {>=}{{$\geq$}}2 {<-}{{$\leftarrow$}}2
          {<=}{{$\leq$}}2 {=>}{{$\Rightarrow$}}2
          {>>}{{>>}}2 {>>=}{{>>=}}2
          {|}{{$\mid$}}1}
\makeatletter
\@ifundefined{lhs2tex.lhs2tex.sty.read}%
  {\@namedef{lhs2tex.lhs2tex.sty.read}{}%
   \newcommand\SkipToFmtEnd{}%
   \newcommand\EndFmtInput{}%
   \long\def\SkipToFmtEnd#1\EndFmtInput{}%
  }\SkipToFmtEnd

\newcommand\ReadOnlyOnce[1]{\@ifundefined{#1}{\@namedef{#1}{}}\SkipToFmtEnd}
\usepackage{amstext}
\usepackage{amssymb}
\usepackage{stmaryrd}
\DeclareFontFamily{OT1}{cmtex}{}
\DeclareFontShape{OT1}{cmtex}{m}{n}
  {<5><6><7><8>cmtex8
   <9>cmtex9
   <10><10.95><12><14.4><17.28><20.74><24.88>cmtex10}{}
\DeclareFontShape{OT1}{cmtex}{m}{it}
  {<-> ssub * cmtt/m/it}{}

\DeclareFontShape{OT1}{cmtt}{bx}{n}
  {<5><6><7><8>cmtt8
   <9>cmbtt9
   <10><10.95><12><14.4><17.28><20.74><24.88>cmbtt10}{}
\DeclareFontShape{OT1}{cmtex}{bx}{n}
  {<-> ssub * cmtt/bx/n}{}
\newcommand{\Conid}[1]{\mathit{#1}}
\newcommand{\Varid}[1]{\mathit{#1}}
\newcommand{\anonymous}{\kern0.06em \vbox{\hrule\@width.5em}}
\newcommand{\plus}{\mathbin{+\!\!\!+}}
\newcommand{\bind}{\mathbin{>\!\!\!>\mkern-6.7mu=}}
\renewcommand{\leq}{\leqslant}
\renewcommand{\geq}{\geqslant}
\usepackage{polytable}

\@ifundefined{mathindent}%
  {\newdimen\mathindent\mathindent\leftmargini}%
  {}%

\def\resethooks{%
  \global\let\SaveRestoreHook\empty
  \global\let\ColumnHook\empty}
\newcommand*{\savecolumns}[1][default]%
  {\g@addto@macro\SaveRestoreHook{\savecolumns[#1]}}
\newcommand*{\restorecolumns}[1][default]%
  {\g@addto@macro\SaveRestoreHook{\restorecolumns[#1]}}
\newcommand*{\aligncolumn}[2]%
  {\g@addto@macro\ColumnHook{\column{#1}{#2}}}

\resethooks

\newcommand{\onelinecommentchars}{\quad-{}- }
\newcommand{\commentbeginchars}{\enskip\{-}
\newcommand{\commentendchars}{-\}\enskip}

\newcommand{\visiblecomments}{%
  \let\onelinecomment=\onelinecommentchars
  \let\commentbegin=\commentbeginchars
  \let\commentend=\commentendchars}

\newcommand{\invisiblecomments}{%
  \let\onelinecomment=\empty
  \let\commentbegin=\empty
  \let\commentend=\empty}

\visiblecomments

\newlength{\blanklineskip}
\newcommand{\hsindent}[1]{\quad}% default is fixed indentation
\let\hspre\empty
\let\hspost\empty

\EndFmtInput
\makeatother
\ReadOnlyOnce{polycode.fmt}%
\makeatletter

\newcommand{\hsnewpar}[1]%
  {{\parskip=0pt\parindent=0pt\par\vskip #1\noindent}}

\newcommand{\hscodestyle}{}

\newcommand{\sethscode}[1]%
  {\expandafter\let\expandafter\hscode\csname #1\endcsname
   \expandafter\let\expandafter\endhscode\csname end#1\endcsname}

  {\par\noindent
   \advance\leftskip\mathindent
   \hscodestyle
   \let\\=\@normalcr
   \let\hspre\(\let\hspost\)%
   \pboxed}%
  {\endpboxed\)%
   \par\noindent
   \ignorespacesafterend}

  {\hsnewpar\abovedisplayskip
   \advance\leftskip\mathindent
   \hscodestyle
   \let\hspre\(\let\hspost\)%
   \pboxed}%
  {\endpboxed%
   \hsnewpar\belowdisplayskip
   \ignorespacesafterend}

  {\hsnewpar\abovedisplayskip
   \advance\leftskip\mathindent
   \hscodestyle
   \let\\=\@normalcr
   \(\pboxed}%
  {\endpboxed\)%
   \hsnewpar\belowdisplayskip
   \ignorespacesafterend}

\newcommand{\plainhs}{\sethscode{plainhscode}}

\plainhs

  {\hsnewpar\abovedisplayskip
   \advance\leftskip\mathindent
   \hscodestyle
   \let\\=\@normalcr
   \(\parray}%
  {\endparray\)%
   \hsnewpar\belowdisplayskip
   \ignorespacesafterend}

  {\parray}{\endparray}

  {\(\parray}{\endparray\)}

\def\codeframewidth{\arrayrulewidth}
\RequirePackage{calc}

  {\parskip=\abovedisplayskip\par\noindent
   \hscodestyle
   \arrayrulewidth=\codeframewidth
   \tabular{@{}|p{\linewidth-2\arraycolsep-2\arrayrulewidth-2pt}|@{}}%
   \hline\framedhslinecorrect\\{-1.5ex}%
   \let\endoflinesave=\\
   \let\\=\@normalcr
   \(\pboxed}%
  {\endpboxed\)%
   \framedhslinecorrect\endoflinesave{.5ex}\hline
   \endtabular
   \parskip=\belowdisplayskip\par\noindent
   \ignorespacesafterend}

\newcommand{\framedhslinecorrect}[2]%
  {#1[#2]}

  {\(\def\column##1##2{}%
   \let\>\undefined\let\<\undefined\let\\\undefined
   \newcommand\>[1][]{}\newcommand\<[1][]{}\newcommand\\[1][]{}%
   \def\fromto##1##2##3{##3}%
   }{\) }%

  {\let\orighscode=\hscode
   \let\origendhscode=\endhscode
   \def\endhscode{\def\hscode{\endgroup\def\@currenvir{hscode}\\}\begingroup}
   \orighscode\def\hscode{\endgroup\def\@currenvir{hscode}}}%
  {\origendhscode
   \global\let\hscode=\orighscode
   \global\let\endhscode=\origendhscode}%

\makeatother
\EndFmtInput

\title{Parametric Compositional Data Types}
\author{Patrick~Bahr \qquad\qquad Tom~Hvitved
  \institute{Department of Computer Science, University of
    Copenhagen\\ Universitetsparken 1, 2100 Copenhagen, Denmark
    \email{\mbox{$\{$}paba,hvitved\mbox{$\}$}@diku.dk}}}
\def\titlerunning{Parametric Compositional Data Types}
\def\authorrunning{P. Bahr \& T. Hvitved}

\begin{document}

% Decrease spacing before and after code blocks
\setlength{\abovedisplayskip}{6pt}
\setlength{\belowdisplayskip}{6pt}

\maketitle

\begin{abstract}
  In previous work we have illustrated the benefits that
  \emph{compositional data types} (CDTs) offer for implementing
  languages and in general for dealing with abstract syntax trees
  (ASTs). Based on Swierstra's \emph{data types \`{a} la carte}, CDTs
  are implemented as a Haskell library that enables the definition of
  recursive data types and functions on them in a modular and
  extendable fashion. Although CDTs provide a powerful tool for
  analysing and manipulating ASTs, they lack a convenient
  representation of variable binders. In this paper we remedy this
  deficiency by combining the framework of CDTs with Chlipala's
  parametric higher-order abstract syntax (PHOAS). We show how a
  generalisation from functors to difunctors enables us to capture
  PHOAS while still maintaining the features of the original
  implementation of CDTs, in particular its modularity. Unlike
  previous approaches, we avoid so-called \emph{exotic terms} without
  resorting to abstract types: this is crucial when we want to perform
  \emph{transformations} on CDTs that inspect the recursively computed
  CDTs, e.g.\ constant folding.
\end{abstract}

\section{Introduction}
\label{sec:introduction}

When implementing domain-specific languages (DSLs)---either as
embedded languages or stand-alone languages---the abstract syntax
trees (ASTs) of programs are usually represented as elements of a
recursive algebraic data type. These ASTs typically undergo various
transformation steps, such as desugaring from a full language to a
core language. But reflecting the invariants of these transformations
in the type system of the host language can be problematic. For
instance, in order to reflect a desugaring transformation in the type
system, we must define a separate data type for ASTs of
the core language. Unfortunately, this has the side effect that common
functionality, such as pretty printing, has to be duplicated.

Wadler identified the essence of this issue as the \emph{Expression
  Problem}, i.e.\ ``the goal [\dots] to define a datatype by cases,
where one can add new cases to the datatype and new functions over the
datatype, without recompiling existing code, and while retaining
static type
safety''~\cite{wadler98exp}. Swierstra~\cite{swierstra08jfp} elegantly
addressed this problem using Haskell and its type classes
machinery. While Swierstra's approach exhibits invaluable simplicity
and clarity, it lacks features necessary to apply it in a practical
setting beyond the confined simplicity of the expression problem. To
this end, the framework of \emph{compositional data types}
(CDTs)~\cite{bahr11wgp} provides a rich library for implementing
practical functionality on highly modular data types. This includes
support of a wide array of recursion schemes in both pure and monadic
forms, as well as mutually recursive data types and generalised
algebraic data types (GADTs)~\cite{schrijvers09icfp}.

What CDTs fail to address, however, is a transparent representation of
variable binders that frees the programmer's mind from common issues
like computations modulo $\alpha$-equivalence and capture-avoiding
substitutions. The work we present in this paper fills that gap by
adopting (a restricted form of) higher-order abstract syntax
(HOAS)~\cite{pfenning88pldi}, which uses the host language's variable
binding mechanism to represent binders in the object language. Since
implementing efficient recursion schemes in the presence of HOAS is
challenging~\cite{fegaras96popl,meijer95fpca,schuermann01tcs,washburn08jfp},
integrating this technique with CDTs is a non-trivial task.

Following a brief introduction to CDTs in
Section~\ref{sec:compositional-data-types}, we describe how to achieve
this integration as follows:
\begin{itemize}
\item We adopt parametric higher-order abstract syntax
  (PHOAS)~\cite{chlipala08icfp}, and we show how to capture this
  restricted form of HOAS via difunctors. The thus obtained
  \emph{parametric compositional data types} (PCDTs) allow for the
  definition of modular catamorphisms \`{a} la Fegaras and
  Sheard~\cite{fegaras96popl} in the presence of binders. Unlike
  previous approaches, our technique does not rely on abstract types,
  which is crucial for modular computations that are also modular in
  their result type
  (Section~\ref{sec:parametric-compositional-data-types}).
\item We illustrate why monadic computations constitute a challenge
  in the parametric setting and we show how monadic
  catamorphisms can still be defined for a restricted class of PCDTs
  (Section~\ref{sec:monadic-computations}).
 \item We show how to transfer the restricted recursion scheme of
   \emph{term homomorphisms}~\cite{bahr11wgp} to PCDTs. Term
   homomorphisms enable the same flexibility for reuse and opportunity
   for deforestation~\cite{wadler90tcs} that we know from CDTs
   (Section~\ref{sec:contexts-and-term-homomorphisms}).
 \item We show how to represent mutually recursive data types and GADTs
  by generalising PCDTs in the style of Johann and
  Ghani~\cite{johann08popl}
  (Section~\ref{sec:generalised-comp-data-types}).
\item We illustrate the practical applicability of our framework by
  means of a complete library example, and we show how to
  automatically derive functionality for deciding equality
  (Section~\ref{sec:pract-cons}).
\end{itemize}

Parametric compositional data types are available as a Haskell
library\footnote{See
  \url{http://hackage.haskell.org/package/compdata}.}, including
numerous examples that are not included in this paper. All code
fragments presented throughout the paper are written in (literate)
Haskell~\cite{marlow10haskell}, and the library relies on several
language extensions that are currently only known to be supported by
the Glasgow Haskell Compiler (GHC).

\section{Compositional Data Types}
\label{sec:compositional-data-types}

Based on Swierstra's \dalc~\cite{swierstra08jfp}, compositional data
types (CDTs)~\cite{bahr11wgp} provide a framework for
manipulating recursive data structures in a type-safe, modular
manner. The prime application of CDTs is within language
implementation and AST manipulation, and we present the basic concepts
of CDTs in this section. More advanced concepts
are introduced in Sections~\ref{sec:monadic-computations},
\ref{sec:contexts-and-term-homomorphisms}, and
\ref{sec:generalised-comp-data-types}.

\subsection{Motivating Example}
\label{sec:motivating-example}

Consider an extension of the lambda calculus with integers, addition,
let expressions, and error signalling:
\[
e ::= \lambda x. e \mid x \mid e_1\: e_2 \mid n \mid e_1 + e_2 \mid
\textbf{let } x=e_1 \textbf{ in } e_2 \mid \textbf{error}
\]
Our goal is to implement a pretty printer, a desugaring
transformation, constant folding, and a call-by-value interpreter for
the simple language above. The desugaring transformation will turn let
expressions $\textbf{let } x=e_1 \textbf{ in } e_2$ into $(\lambda
x. e_2)\; e_1$. Constant folding and evaluation will take place after
desugaring, i.e.\ both computations are only defined for the core
language without let expressions.

The standard approach to representing the language above is in terms
of an algebraic data type:
\begin{hscode}\SaveRestoreHook
\column{B}{@{}>{\hspre}l<{\hspost}@{}}%
\column{11}{@{}>{\hspre}l<{\hspost}@{}}%
\column{E}{@{}>{\hspre}l<{\hspost}@{}}%
\>[B]{}\mathbf{type}\;\Conid{Var}{}\<[11]%
\>[11]{}\mathrel{=}\Conid{String}{}\<[E]%
\\[\blanklineskip]%
\>[B]{}\mathbf{data}\;\Conid{Exp}{}\<[11]%
\>[11]{}\mathrel{=}\Conid{Lam}\;\Conid{Var}\;\Conid{Exp}\mid \Conid{Var}\;\Conid{Var}\mid \Conid{App}\;\Conid{Exp}\;\Conid{Exp}\mid \Conid{Lit}\;\Conid{Int}\mid \Conid{Plus}\;\Conid{Exp}\;\Conid{Exp}\mid \Conid{Let}\;\Conid{Var}\;\Conid{Exp}\;\Conid{Exp}\mid \Conid{Err}{}\<[E]%
\ColumnHook
\end{hscode}\resethooks
We may then straightforwardly define the pretty printer \ensuremath{\Varid{pretty}\mathbin{::}\Conid{Exp}\to \Conid{String}}. However, when we want to implement the desugaring
transformation, we need a new algebraic data type:
\begin{hscode}\SaveRestoreHook
\column{B}{@{}>{\hspre}l<{\hspost}@{}}%
\column{E}{@{}>{\hspre}l<{\hspost}@{}}%
\>[B]{}\mathbf{data}\;\Conid{Exp'}\mathrel{=}\Conid{Lam'}\;\Conid{Var}\;\Conid{Exp'}\mid \Conid{Var'}\;\Conid{Var}\mid \Conid{App'}\;\Conid{Exp'}\;\Conid{Exp'}\mid \Conid{Lit'}\;\Conid{Int}\mid \Conid{Plus'}\;\Conid{Exp'}\;\Conid{Exp'}\mid \Conid{Err'}{}\<[E]%
\ColumnHook
\end{hscode}\resethooks
That is, we need to replicate all constructors of \ensuremath{\Conid{Exp}}---except
\ensuremath{\Conid{Let}}---into a new type \ensuremath{\Conid{Exp'}} of core expressions, in order to obtain
a properly typed desugaring function \ensuremath{\Varid{desug}\mathbin{::}\Conid{Exp}\to \Conid{Exp'}}. Not only
does this mean that we have to replicate the constructors, we also
need to replicate common functionality, e.g.\ in order to obtain a
pretty printer for \ensuremath{\Conid{Exp'}} we must either write a new function, or
write an injection function \ensuremath{\Conid{Exp'}\to \Conid{Exp}}.

CDTs provide a solution that allows us to define the ASTs for (core)
expressions without having to duplicate common constructors, and
without having to give up on statically guaranteed invariants about
the structure of the ASTs. CDTs take the viewpoint of data types as
fixed points of functors~\cite{meijer91fpca}, i.e.\ the definition of
the AST data type is separated into non-recursive signatures
(functors) on the one hand and the recursive structure on the other
hand. For our example, we define the following signatures (omitting
the straightforward \ensuremath{\Conid{Functor}} instance declarations):
\begin{hscode}\SaveRestoreHook
\column{B}{@{}>{\hspre}l<{\hspost}@{}}%
\column{15}{@{}>{\hspre}c<{\hspost}@{}}%
\column{15E}{@{}l@{}}%
\column{18}{@{}>{\hspre}l<{\hspost}@{}}%
\column{44}{@{}>{\hspre}l<{\hspost}@{}}%
\column{57}{@{}>{\hspre}c<{\hspost}@{}}%
\column{57E}{@{}l@{}}%
\column{60}{@{}>{\hspre}l<{\hspost}@{}}%
\column{85}{@{}>{\hspre}l<{\hspost}@{}}%
\column{97}{@{}>{\hspre}c<{\hspost}@{}}%
\column{97E}{@{}l@{}}%
\column{100}{@{}>{\hspre}l<{\hspost}@{}}%
\column{E}{@{}>{\hspre}l<{\hspost}@{}}%
\>[B]{}\mathbf{data}\;\Conid{Lam}\;\Varid{a}{}\<[15]%
\>[15]{}\mathrel{=}{}\<[15E]%
\>[18]{}\Conid{Lam}\;\Conid{Var}\;\Varid{a}$\qquad$\;{}\<[44]%
\>[44]{}\mathbf{data}\;\Conid{Lit}\;\Varid{a}{}\<[57]%
\>[57]{}\mathrel{=}{}\<[57E]%
\>[60]{}\Conid{Lit}\;\Conid{Int}\;{}\<[85]%
\>[85]{}\mathbf{data}\;\Conid{Let}\;\Varid{a}{}\<[97]%
\>[97]{}\mathrel{=}{}\<[97E]%
\>[100]{}\Conid{Let}\;\Conid{Var}\;\Varid{a}\;\Varid{a}{}\<[E]%
\\[\blanklineskip]%
\>[B]{}\mathbf{data}\;\Conid{Var}\;\Varid{a}{}\<[15]%
\>[15]{}\mathrel{=}{}\<[15E]%
\>[18]{}\Conid{Var}\;\Conid{Var}\;{}\<[44]%
\>[44]{}\mathbf{data}\;\Conid{Plus}\;\Varid{a}{}\<[57]%
\>[57]{}\mathrel{=}{}\<[57E]%
\>[60]{}\Conid{Plus}\;\Varid{a}\;\Varid{a}$\qquad$\;{}\<[85]%
\>[85]{}\mathbf{data}\;\Conid{Err}\;\Varid{a}{}\<[97]%
\>[97]{}\mathrel{=}{}\<[97E]%
\>[100]{}\Conid{Err}{}\<[E]%
\\[\blanklineskip]%
\>[B]{}\mathbf{data}\;\Conid{App}\;\Varid{a}{}\<[15]%
\>[15]{}\mathrel{=}{}\<[15E]%
\>[18]{}\Conid{App}\;\Varid{a}\;\Varid{a}{}\<[E]%
\ColumnHook
\end{hscode}\resethooks
Signatures can then be combined in a modular fashion by means of a
formal sum of functors:
\begin{hscode}\SaveRestoreHook
\column{B}{@{}>{\hspre}l<{\hspost}@{}}%
\column{3}{@{}>{\hspre}l<{\hspost}@{}}%
\column{12}{@{}>{\hspre}l<{\hspost}@{}}%
\column{19}{@{}>{\hspre}l<{\hspost}@{}}%
\column{E}{@{}>{\hspre}l<{\hspost}@{}}%
\>[B]{}\mathbf{data}\;(\Varid{f}\fplus\Varid{g})\;\Varid{a}\mathrel{=}\Conid{Inl}\;(\Varid{f}\;\Varid{a})\mid \Conid{Inr}\;(\Varid{g}\;\Varid{a}){}\<[E]%
\\[\blanklineskip]%
\>[B]{}\mathbf{instance}\;(\Conid{Functor}\;\Varid{f},\Conid{Functor}\;\Varid{g})\Rightarrow \Conid{Functor}\;(\Varid{f}\fplus\Varid{g})\;\mathbf{where}{}\<[E]%
\\
\>[B]{}\hsindent{3}{}\<[3]%
\>[3]{}\Varid{fmap}\;\Varid{f}\;(\Conid{Inl}\;\Varid{x}){}\<[19]%
\>[19]{}\mathrel{=}\Conid{Inl}\;(\Varid{fmap}\;\Varid{f}\;\Varid{x}){}\<[E]%
\\
\>[B]{}\hsindent{3}{}\<[3]%
\>[3]{}\Varid{fmap}\;\Varid{f}\;(\Conid{Inr}\;\Varid{x}){}\<[19]%
\>[19]{}\mathrel{=}\Conid{Inr}\;(\Varid{fmap}\;\Varid{f}\;\Varid{x}){}\<[E]%
\\[\blanklineskip]%
\>[B]{}\mathbf{type}\;\Conid{Sig}{}\<[12]%
\>[12]{}\mathrel{=}\Conid{Lam}\fplus\Conid{Var}\fplus\Conid{App}\fplus\Conid{Lit}\fplus\Conid{Plus}\fplus\Conid{Err}\fplus\Conid{Let}{}\<[E]%
\\[\blanklineskip]%
\>[B]{}\mathbf{type}\;\Conid{Sig'}{}\<[12]%
\>[12]{}\mathrel{=}\Conid{Lam}\fplus\Conid{Var}\fplus\Conid{App}\fplus\Conid{Lit}\fplus\Conid{Plus}\fplus\Conid{Err}{}\<[E]%
\ColumnHook
\end{hscode}\resethooks

Finally, the type of terms over a (potentially compound) signature \ensuremath{\Varid{f}}
can be constructed as the (least) fixed point of the signature \ensuremath{\Varid{f}}:
\begin{hscode}\SaveRestoreHook
\column{B}{@{}>{\hspre}l<{\hspost}@{}}%
\column{E}{@{}>{\hspre}l<{\hspost}@{}}%
\>[B]{}\mathbf{data}\;\Conid{Term}\;\Varid{f}\mathrel{=}\Conid{In}\;\{\mskip1.5mu \Varid{out}\mathbin{::}\Varid{f}\;(\Conid{Term}\;\Varid{f})\mskip1.5mu\}{}\<[E]%
\ColumnHook
\end{hscode}\resethooks
Modulo strictness, \ensuremath{\Conid{Term}\;\Conid{Sig}} is isomorphic to \ensuremath{\Conid{Exp}}, and \ensuremath{\Conid{Term}\;\Conid{Sig'}}
is isomorphic to \ensuremath{\Conid{Exp'}}.

The use of formal sums entails that each (sub)term has to be
explicitly tagged with zero or more \ensuremath{\Conid{Inl}} or \ensuremath{\Conid{Inr}} tags. In order to
add the right tags automatically, injections are derived using a type
class:
\begin{hscode}\SaveRestoreHook
\column{B}{@{}>{\hspre}l<{\hspost}@{}}%
\column{3}{@{}>{\hspre}l<{\hspost}@{}}%
\column{9}{@{}>{\hspre}l<{\hspost}@{}}%
\column{E}{@{}>{\hspre}l<{\hspost}@{}}%
\>[B]{}\mathbf{class}\;\Varid{sub}\fsub\Varid{sup}\;\mathbf{where}{}\<[E]%
\\
\>[B]{}\hsindent{3}{}\<[3]%
\>[3]{}\Varid{inj}{}\<[9]%
\>[9]{}\mathbin{::}\Varid{sub}\;\Varid{a}\to \Varid{sup}\;\Varid{a}{}\<[E]%
\\
\>[B]{}\hsindent{3}{}\<[3]%
\>[3]{}\Varid{proj}{}\<[9]%
\>[9]{}\mathbin{::}\Varid{sup}\;\Varid{a}\to \Conid{Maybe}\;(\Varid{sub}\;\Varid{a}){}\<[E]%
\ColumnHook
\end{hscode}\resethooks
Using \emph{overlapping instance} declarations, the subsignature
relation \ensuremath{\fsub} can be constructively
defined~\cite{swierstra08jfp}. However, due to the limitations of
Haskell's type class system, instances are restricted to the form \ensuremath{\Varid{f}\fsub\Varid{g}} where \ensuremath{\Varid{f}} is atomic, i.e.\ not a sum, and \ensuremath{\Varid{g}} is a
right-associated sum, e.g.\ \ensuremath{\Varid{g}_{1}\fplus(\Varid{g}_{2}\fplus\Varid{g}_{3})} but not \ensuremath{(\Varid{g}_{1}\fplus\Varid{g}_{2})\fplus\Varid{g}_{3}}. With the carefully defined instances for \ensuremath{\fsub}, injection and
projection functions for terms can then be defined as follows:
\begin{hscode}\SaveRestoreHook
\column{B}{@{}>{\hspre}l<{\hspost}@{}}%
\column{E}{@{}>{\hspre}l<{\hspost}@{}}%
\>[B]{}\Varid{inject}\mathbin{::}(\Varid{g}\fsub\Varid{f})\Rightarrow \Varid{g}\;(\Conid{Term}\;\Varid{f})\to \Conid{Term}\;\Varid{f}{}\<[E]%
\\
\>[B]{}\Varid{inject}\mathrel{=}\Conid{In}\mathbin{.}\Varid{inj}{}\<[E]%
\\[\blanklineskip]%
\>[B]{}\Varid{project}\mathbin{::}(\Varid{g}\fsub\Varid{f})\Rightarrow \Conid{Term}\;\Varid{f}\to \Conid{Maybe}\;(\Varid{g}\;(\Conid{Term}\;\Varid{f})){}\<[E]%
\\
\>[B]{}\Varid{project}\mathrel{=}\Varid{proj}\mathbin{.}\Varid{out}{}\<[E]%
\ColumnHook
\end{hscode}\resethooks

Additionally, in order to reduce the syntactic overhead, the CDTs
library can automatically derive smart constructors that comprise the
injections~\cite{bahr11wgp}, e.g.\
\begin{hscode}\SaveRestoreHook
\column{B}{@{}>{\hspre}l<{\hspost}@{}}%
\column{E}{@{}>{\hspre}l<{\hspost}@{}}%
\>[B]{}\Varid{iPlus}\mathbin{::}(\Conid{Plus}\fsub\Varid{f})\Rightarrow \Conid{Term}\;\Varid{f}\to \Conid{Term}\;\Varid{f}\to \Conid{Term}\;\Varid{f}{}\<[E]%
\\
\>[B]{}\Varid{iPlus}\;\Varid{x}\;\Varid{y}\mathrel{=}\Varid{inject}\;(\Conid{Plus}\;\Varid{x}\;\Varid{y}){}\<[E]%
\ColumnHook
\end{hscode}\resethooks
Using the derived smart constructors, we can then write expressions
such as $\textbf{let } x = 2 \textbf{ in } (\lambda y. y + x)\; 3$
without syntactic overhead:
\begin{hscode}\SaveRestoreHook
\column{B}{@{}>{\hspre}l<{\hspost}@{}}%
\column{E}{@{}>{\hspre}l<{\hspost}@{}}%
\>[B]{}\Varid{e}\mathbin{::}\Conid{Term}\;\Conid{Sig}{}\<[E]%
\\
\>[B]{}\Varid{e}\mathrel{=}\Varid{iLet}\;\text{\tt \char34 x\char34}\;(\Varid{iLit}\;\mathrm{2})\;((\Varid{iLam}\;\text{\tt \char34 y\char34}\;(\Conid{Var}\;\text{\tt \char34 y\char34}\mathbin{`\Varid{iPlus}`}\Conid{Var}\;\text{\tt \char34 x\char34}))\mathbin{`\Varid{iApp}`}\Varid{iLit}\;\mathrm{3}){}\<[E]%
\ColumnHook
\end{hscode}\resethooks
In fact, the principal type of \ensuremath{\Varid{e}} is the \emph{open} type:
\begin{hscode}\SaveRestoreHook
\column{B}{@{}>{\hspre}l<{\hspost}@{}}%
\column{E}{@{}>{\hspre}l<{\hspost}@{}}%
\>[B]{}(\Conid{Lam}\fsub\Varid{f},\Conid{Var}\fsub\Varid{f},\Conid{App}\fsub\Varid{f},\Conid{Lit}\fsub\Varid{f},\Conid{Plus}\fsub\Varid{f},\Conid{Let}\fsub\Varid{f})\Rightarrow \Conid{Term}\;\Varid{f}{}\<[E]%
\ColumnHook
\end{hscode}\resethooks
which means that \ensuremath{\Varid{e}} can be used as a term over any signature
containing at least these six signatures!

Next, we want to define the pretty printer, i.e.\ a function of
type \ensuremath{\Conid{Term}\;\Conid{Sig}\to \Conid{String}}. In order to make a recursive function
definition modular too, it is defined
as the catamorphism of an algebra~\cite{meijer91fpca}:
\begin{hscode}\SaveRestoreHook
\column{B}{@{}>{\hspre}l<{\hspost}@{}}%
\column{E}{@{}>{\hspre}l<{\hspost}@{}}%
\>[B]{}\mathbf{type}\;\Conid{Alg}\;\Varid{f}\;\Varid{a}\mathrel{=}\Varid{f}\;\Varid{a}\to \Varid{a}{}\<[E]%
\\[\blanklineskip]%
\>[B]{}\Varid{cata}\mathbin{::}\Conid{Functor}\;\Varid{f}\Rightarrow \Conid{Alg}\;\Varid{f}\;\Varid{a}\to \Conid{Term}\;\Varid{f}\to \Varid{a}{}\<[E]%
\\
\>[B]{}\Varid{cata}\;\phi\mathrel{=}\phi\mathbin{.}\Varid{fmap}\;(\Varid{cata}\;\phi)\mathbin{.}\Varid{out}{}\<[E]%
\ColumnHook
\end{hscode}\resethooks
The advantage of this approach is that algebras can be easily combined
over formal sums. A modular algebra definition is obtained by an open
family of algebras indexed by the signature and closed under forming
formal sums. This is achieved as a type class:
\begin{hscode}\SaveRestoreHook
\column{B}{@{}>{\hspre}l<{\hspost}@{}}%
\column{3}{@{}>{\hspre}l<{\hspost}@{}}%
\column{22}{@{}>{\hspre}c<{\hspost}@{}}%
\column{22E}{@{}l@{}}%
\column{25}{@{}>{\hspre}l<{\hspost}@{}}%
\column{E}{@{}>{\hspre}l<{\hspost}@{}}%
\>[B]{}\mathbf{class}\;\Conid{Pretty}\;\Varid{f}\;\mathbf{where}{}\<[E]%
\\
\>[B]{}\hsindent{3}{}\<[3]%
\>[3]{}\phi_{\mathrm{Pretty}}\mathbin{::}\Conid{Alg}\;\Varid{f}\;\Conid{String}{}\<[E]%
\\[\blanklineskip]%
\>[B]{}\mathbf{instance}\;(\Conid{Pretty}\;\Varid{f},\Conid{Pretty}\;\Varid{g})\Rightarrow \Conid{Pretty}\;(\Varid{f}\fplus\Varid{g})\;\mathbf{where}{}\<[E]%
\\
\>[B]{}\hsindent{3}{}\<[3]%
\>[3]{}\phi_{\mathrm{Pretty}}\;(\Conid{Inl}\;\Varid{x}){}\<[22]%
\>[22]{}\mathrel{=}{}\<[22E]%
\>[25]{}\phi_{\mathrm{Pretty}}\;\Varid{x}{}\<[E]%
\\
\>[B]{}\hsindent{3}{}\<[3]%
\>[3]{}\phi_{\mathrm{Pretty}}\;(\Conid{Inr}\;\Varid{x}){}\<[22]%
\>[22]{}\mathrel{=}{}\<[22E]%
\>[25]{}\phi_{\mathrm{Pretty}}\;\Varid{x}{}\<[E]%
\\[\blanklineskip]%
\>[B]{}\Varid{pretty}\mathbin{::}(\Conid{Functor}\;\Varid{f},\Conid{Pretty}\;\Varid{f})\Rightarrow \Conid{Term}\;\Varid{f}\to \Conid{String}{}\<[E]%
\\
\>[B]{}\Varid{pretty}\mathrel{=}\Varid{cata}\;\phi_{\mathrm{Pretty}}{}\<[E]%
\ColumnHook
\end{hscode}\resethooks

The instance declaration that lifts \ensuremath{\Conid{Pretty}} instances to sums is
crucial. Yet, the structure of its declaration is independent from the
particular algebra class, and the CDTs library provides a mechanism for
automatically deriving such instances~\cite{bahr11wgp}.
What remains in order to implement the pretty printer is to define
instances of the \ensuremath{\Conid{Pretty}} algebra class for the six signatures:
\begin{hscode}\SaveRestoreHook
\column{B}{@{}>{\hspre}l<{\hspost}@{}}%
\column{3}{@{}>{\hspre}l<{\hspost}@{}}%
\column{E}{@{}>{\hspre}l<{\hspost}@{}}%
\>[B]{}\mathbf{instance}\;\Conid{Pretty}\;\Conid{Lam}\;\mathbf{where}{}\<[E]%
\\
\>[B]{}\hsindent{3}{}\<[3]%
\>[3]{}\phi_{\mathrm{Pretty}}\;(\Conid{Lam}\;\Varid{x}\;\Varid{e})\mathrel{=}\text{\tt \char34 (\char92 \char92 \char34}\plus \Varid{x}\plus \text{\tt \char34 .~\char34}\plus \Varid{e}\plus \text{\tt \char34 )\char34}{}\<[E]%
\\[\blanklineskip]%
\>[B]{}\mathbf{instance}\;\Conid{Pretty}\;\Conid{Var}\;\mathbf{where}{}\<[E]%
\\
\>[B]{}\hsindent{3}{}\<[3]%
\>[3]{}\phi_{\mathrm{Pretty}}\;(\Conid{Var}\;\Varid{x})\mathrel{=}\Varid{x}{}\<[E]%
\\[\blanklineskip]%
\>[B]{}\mathbf{instance}\;\Conid{Pretty}\;\Conid{App}\;\mathbf{where}{}\<[E]%
\\
\>[B]{}\hsindent{3}{}\<[3]%
\>[3]{}\phi_{\mathrm{Pretty}}\;(\Conid{App}\;\Varid{e}_{1}\;\Varid{e}_{2})\mathrel{=}\text{\tt \char34 (\char34}\plus \Varid{e}_{1}\plus \text{\tt \char34 ~\char34}\plus \Varid{e}_{2}\plus \text{\tt \char34 )\char34}{}\<[E]%
\\[\blanklineskip]%
\>[B]{}\mathbf{instance}\;\Conid{Pretty}\;\Conid{Lit}\;\mathbf{where}{}\<[E]%
\\
\>[B]{}\hsindent{3}{}\<[3]%
\>[3]{}\phi_{\mathrm{Pretty}}\;(\Conid{Lit}\;\Varid{n})\mathrel{=}\Varid{show}\;\Varid{n}{}\<[E]%
\\[\blanklineskip]%
\>[B]{}\mathbf{instance}\;\Conid{Pretty}\;\Conid{Plus}\;\mathbf{where}{}\<[E]%
\\
\>[B]{}\hsindent{3}{}\<[3]%
\>[3]{}\phi_{\mathrm{Pretty}}\;(\Conid{Plus}\;\Varid{e}_{1}\;\Varid{e}_{2})\mathrel{=}\text{\tt \char34 (\char34}\plus \Varid{e}_{1}\plus \text{\tt \char34 ~+~\char34}\plus \Varid{e}_{2}\plus \text{\tt \char34 )\char34}{}\<[E]%
\\[\blanklineskip]%
\>[B]{}\mathbf{instance}\;\Conid{Pretty}\;\Conid{Let}\;\mathbf{where}{}\<[E]%
\\
\>[B]{}\hsindent{3}{}\<[3]%
\>[3]{}\phi_{\mathrm{Pretty}}\;(\Conid{Let}\;\Varid{x}\;\Varid{e}_{1}\;\Varid{e}_{2})\mathrel{=}\text{\tt \char34 (let~\char34}\plus \Varid{x}\plus \text{\tt \char34 ~=~\char34}\plus \Varid{e}_{1}\plus \text{\tt \char34 ~in~\char34}\plus \Varid{e}_{2}\plus \text{\tt \char34 )\char34}{}\<[E]%
\\[\blanklineskip]%
\>[B]{}\mathbf{instance}\;\Conid{Pretty}\;\Conid{Err}\;\mathbf{where}{}\<[E]%
\\
\>[B]{}\hsindent{3}{}\<[3]%
\>[3]{}\phi_{\mathrm{Pretty}}\;\Conid{Err}\mathrel{=}\text{\tt \char34 error\char34}{}\<[E]%
\ColumnHook
\end{hscode}\resethooks
With these definitions we then have that \ensuremath{\Varid{pretty}\;\Varid{e}} evaluates to the
string \texttt{(let x = 2 in  ((\char`\\ y.\;(y + x)) 3))}. Moreover, we
automatically obtain a pretty printer for the core language as well,
cf.\ the type of \ensuremath{\Varid{pretty}}.

\section{Parametric Compositional Data Types}
\label{sec:parametric-compositional-data-types}

In the previous section we considered a first-order encoding of the
language, which means that we have to be careful to ensure that
computations are invariant under $\alpha$-equivalence, e.g.\ when
implementing capture-avoiding substitutions. \emph{Higher-order
  abstract syntax} (HOAS)~\cite{pfenning88pldi} remedies this issue,
by representing binders and variables of the object language in terms
of those of the meta language.

\subsection{Higher-Order Abstract Syntax}
\label{sec:hoas}

In a standard Haskell HOAS encoding we replace the signatures \ensuremath{\Conid{Var}}
and \ensuremath{\Conid{Lam}} by a revised \ensuremath{\Conid{Lam}} signature:
\begin{hscode}\SaveRestoreHook
\column{B}{@{}>{\hspre}l<{\hspost}@{}}%
\column{E}{@{}>{\hspre}l<{\hspost}@{}}%
\>[B]{}\mathbf{data}\;\Conid{Lam}\;\Varid{a}\mathrel{=}\Conid{Lam}\;(\Varid{a}\to \Varid{a}){}\<[E]%
\ColumnHook
\end{hscode}\resethooks
Now, however, \ensuremath{\Conid{Lam}} is no longer an instance of \ensuremath{\Conid{Functor}}, because \ensuremath{\Varid{a}}
occurs both in a contravariant position and a covariant position. We
therefore need to generalise functors in order to allow for negative
occurrences of the recursive
parameter. \emph{Difunctors}~\cite{meijer95fpca} provide such a
generalisation:
\begin{hscode}\SaveRestoreHook
\column{B}{@{}>{\hspre}l<{\hspost}@{}}%
\column{3}{@{}>{\hspre}l<{\hspost}@{}}%
\column{E}{@{}>{\hspre}l<{\hspost}@{}}%
\>[B]{}\mathbf{class}\;\Conid{Difunctor}\;\Varid{f}\;\mathbf{where}{}\<[E]%
\\
\>[B]{}\hsindent{3}{}\<[3]%
\>[3]{}\Varid{dimap}\mathbin{::}(\Varid{a}\to \Varid{b})\to (\Varid{c}\to \Varid{d})\to \Varid{f}\;\Varid{b}\;\Varid{c}\to \Varid{f}\;\Varid{a}\;\Varid{d}{}\<[E]%
\\[\blanklineskip]%
\>[B]{}\mathbf{instance}\;\Conid{Difunctor}\;(\to )\;\mathbf{where}{}\<[E]%
\\
\>[B]{}\hsindent{3}{}\<[3]%
\>[3]{}\Varid{dimap}\;\Varid{f}\;\Varid{g}\;\Varid{h}\mathrel{=}\Varid{g}\mathbin{.}\Varid{h}\mathbin{.}\Varid{f}{}\<[E]%
\\[\blanklineskip]%
\>[B]{}\mathbf{instance}\;\Conid{Difunctor}\;\Varid{f}\Rightarrow \Conid{Functor}\;(\Varid{f}\;\Varid{a})\;\mathbf{where}{}\<[E]%
\\
\>[B]{}\hsindent{3}{}\<[3]%
\>[3]{}\Varid{fmap}\mathrel{=}\Varid{dimap}\;\Varid{id}{}\<[E]%
\ColumnHook
\end{hscode}\resethooks
A difunctor must preserve the identity function and distribute over
function composition:
\begin{center}
\ensuremath{\Varid{dimap}\;\Varid{id}\;\Varid{id}\mathrel{=}\Varid{id}} \;\;\;\; and \;\;\;\; \ensuremath{\Varid{dimap}\;(\Varid{f}\mathbin{.}\Varid{g})\;(\Varid{h}\mathbin{.}\Varid{i})\mathrel{=}\Varid{dimap}\;\Varid{g}\;\Varid{h}\mathbin{.}\Varid{dimap}\;\Varid{f}\;\Varid{i}}    
\end{center}
The derived \ensuremath{\Conid{Functor}} instance obtained by fixing the contravariant argument will hence satisfy the functor laws,
provided that the difunctor laws are satisfied.

Meijer and Hutton~\cite{meijer95fpca} showed that it is possible to
perform recursion over difunctor terms:
\begin{hscode}\SaveRestoreHook
\column{B}{@{}>{\hspre}l<{\hspost}@{}}%
\column{E}{@{}>{\hspre}l<{\hspost}@{}}%
\>[B]{}\mathbf{data}\;Term_{MH}\;\Varid{f}\mathrel{=}In_{MH}\;\{\mskip1.5mu out_{MH}\mathbin{::}\Varid{f}\;(Term_{MH}\;\Varid{f})\;(Term_{MH}\;\Varid{f})\mskip1.5mu\}{}\<[E]%
\\[\blanklineskip]%
\>[B]{}cata_{MH}\mathbin{::}\Conid{Difunctor}\;\Varid{f}\Rightarrow (\Varid{f}\;\Varid{b}\;\Varid{a}\to \Varid{a})\to (\Varid{b}\to \Varid{f}\;\Varid{a}\;\Varid{b})\to Term_{MH}\;\Varid{f}\to \Varid{a}{}\<[E]%
\\
\>[B]{}cata_{MH}\;\phi\;\psi\mathrel{=}\phi\mathbin{.}\Varid{dimap}\;(ana_{MH}\;\phi\;\psi)\;(cata_{MH}\;\phi\;\psi)\mathbin{.}out_{MH}{}\<[E]%
\\[\blanklineskip]%
\>[B]{}ana_{MH}\mathbin{::}\Conid{Difunctor}\;\Varid{f}\Rightarrow (\Varid{f}\;\Varid{b}\;\Varid{a}\to \Varid{a})\to (\Varid{b}\to \Varid{f}\;\Varid{a}\;\Varid{b})\to \Varid{b}\to Term_{MH}\;\Varid{f}{}\<[E]%
\\
\>[B]{}ana_{MH}\;\phi\;\psi\mathrel{=}In_{MH}\mathbin{.}\Varid{dimap}\;(cata_{MH}\;\phi\;\psi)\;(ana_{MH}\;\phi\;\psi)\mathbin{.}\psi{}\<[E]%
\ColumnHook
\end{hscode}\resethooks
With Meijer and Hutton's approach, however, in order to lift an
algebra \ensuremath{\phi\mathbin{::}\Varid{f}\;\Varid{b}\;\Varid{a}\to \Varid{a}} to a catamorphism, we also need to supply
the \emph{inverse coalgebra} \ensuremath{\psi\mathbin{::}\Varid{b}\to \Varid{f}\;\Varid{b}\;\Varid{a}}. That is, in order to
write a pretty printer we must supply a parser, which is not
feasible---or perhaps even possible---in practice.

Fortunately, Fegaras and Sheard~\cite{fegaras96popl} realised that if
the embedded functions within terms are \emph{parametric}, then the
inverse coalgebra is only used in order to \emph{undo} computations
performed by the algebra, since parametric functions can only ``push
around their arguments'' without examining them.  The solution
proposed by Fegaras and Sheard is to add a \emph{placeholder} to the
structure of terms, which acts as a right-inverse of the
catamorphism:\footnote{Actually, Fegaras and Sheard do not use
  difunctors, but the given definition corresponds to their encoding.}
\begin{hscode}\SaveRestoreHook
\column{B}{@{}>{\hspre}l<{\hspost}@{}}%
\column{23}{@{}>{\hspre}l<{\hspost}@{}}%
\column{E}{@{}>{\hspre}l<{\hspost}@{}}%
\>[B]{}\mathbf{data}\;Term_{FS}\;\Varid{f}\;\Varid{a}\mathrel{=}In_{FS}\;(\Varid{f}\;(Term_{FS}\;\Varid{f}\;\Varid{a})\;(Term_{FS}\;\Varid{f}\;\Varid{a}))\mid \Conid{Place}\;\Varid{a}{}\<[E]%
\\[\blanklineskip]%
\>[B]{}cata_{FS}\mathbin{::}\Conid{Difunctor}\;\Varid{f}\Rightarrow (\Varid{f}\;\Varid{a}\;\Varid{a}\to \Varid{a})\to Term_{FS}\;\Varid{f}\;\Varid{a}\to \Varid{a}{}\<[E]%
\\
\>[B]{}cata_{FS}\;\phi\;(In_{FS}\;\Varid{t}){}\<[23]%
\>[23]{}\mathrel{=}\phi\;(\Varid{dimap}\;\Conid{Place}\;(cata_{FS}\;\phi)\;\Varid{t}){}\<[E]%
\\
\>[B]{}cata_{FS}\;\phi\;(\Conid{Place}\;\Varid{x}){}\<[23]%
\>[23]{}\mathrel{=}\Varid{x}{}\<[E]%
\ColumnHook
\end{hscode}\resethooks
We can then define e.g.\ a signature for lambda terms, and a
function that calculates the number of bound variables occurring in a
term, as follows (the example is adopted from Washburn and
Weirich~\cite{washburn08jfp}):
\begin{hscode}\SaveRestoreHook
\column{B}{@{}>{\hspre}l<{\hspost}@{}}%
\column{16}{@{}>{\hspre}c<{\hspost}@{}}%
\column{16E}{@{}l@{}}%
\column{19}{@{}>{\hspre}l<{\hspost}@{}}%
\column{E}{@{}>{\hspre}l<{\hspost}@{}}%
\>[B]{}\mathbf{data}\;\Conid{T}\;\Varid{a}\;\Varid{b}\mathrel{=}\Conid{Lam}\;(\Varid{a}\to \Varid{b})\mid \Conid{App}\;\Varid{b}\;\Varid{b}\mbox{\onelinecomment  \ensuremath{\Conid{T}} is a difunctor, we omit the instance declaration}{}\<[E]%
\\[\blanklineskip]%
\>[B]{}\phi\mathbin{::}\Conid{T}\;\Conid{Int}\;\Conid{Int}\to \Conid{Int}{}\<[E]%
\\
\>[B]{}\phi\;(\Conid{Lam}\;\Varid{f}){}\<[16]%
\>[16]{}\mathrel{=}{}\<[16E]%
\>[19]{}\Varid{f}\;\mathrm{1}{}\<[E]%
\\
\>[B]{}\phi\;(\Conid{App}\;\Varid{x}\;\Varid{y}){}\<[16]%
\>[16]{}\mathrel{=}{}\<[16E]%
\>[19]{}\Varid{x}\mathbin{+}\Varid{y}{}\<[E]%
\\[\blanklineskip]%
\>[B]{}\Varid{countVar}\mathbin{::}Term_{FS}\;\Conid{T}\;\Conid{Int}\to \Conid{Int}{}\<[E]%
\\
\>[B]{}\Varid{countVar}\mathrel{=}cata_{FS}\;\phi{}\<[E]%
\ColumnHook
\end{hscode}\resethooks

In the \ensuremath{Term_{FS}} encoding above, however, parametricity of the embedded
functions is not guaranteed. More specifically, the type allows for three
kinds of \emph{exotic terms}~\cite{washburn08jfp}, i.e.\ values in the meta
language that do not correspond to terms in the object language:
\begin{hscode}\SaveRestoreHook
\column{B}{@{}>{\hspre}l<{\hspost}@{}}%
\column{29}{@{}>{\hspre}l<{\hspost}@{}}%
\column{57}{@{}>{\hspre}l<{\hspost}@{}}%
\column{E}{@{}>{\hspre}l<{\hspost}@{}}%
\>[B]{}\Varid{badPlace}\mathbin{::}Term_{FS}\;\Conid{T}\;\Conid{Bool}{}\<[E]%
\\
\>[B]{}\Varid{badPlace}\mathrel{=}In_{FS}\;(\Conid{Place}\;\Conid{True}){}\<[E]%
\\[\blanklineskip]%
\>[B]{}\Varid{badCata}\mathbin{::}Term_{FS}\;\Conid{T}\;\Conid{Int}{}\<[E]%
\\
\>[B]{}\Varid{badCata}\mathrel{=}In_{FS}\;(\Conid{Lam}\;(\lambda \Varid{x}\to \mathbf{if}\;\Varid{countVar}\;\Varid{x}\equiv \mathrm{0}\;\mathbf{then}\;\Varid{x}\;\mathbf{else}\;\Conid{Place}\;\mathrm{0})){}\<[E]%
\\[\blanklineskip]%
\>[B]{}\Varid{badCase}\mathbin{::}Term_{FS}\;\Conid{T}\;\Varid{a}{}\<[E]%
\\
\>[B]{}\Varid{badCase}\mathrel{=}In_{FS}\;(\Conid{Lam}\;(\lambda \Varid{x}\to {}\<[29]%
\>[29]{}\mathbf{case}\;\Varid{x}\;\mathbf{of}\;Term_{FS}\;(\Conid{App}\;\anonymous \;\anonymous ){}\<[57]%
\>[57]{}\to Term_{FS}\;(\Conid{App}\;\Varid{x}\;\Varid{x});\anonymous \to \Varid{x})){}\<[E]%
\ColumnHook
\end{hscode}\resethooks
Fegaras and Sheard showed how to avoid exotic terms by means of a
custom type system. Washburn and Weirich~\cite{washburn08jfp} later
showed that exotic terms can be avoided in a Haskell encoding via type
parametricity and an abstract type of terms: terms are restricted to
the type \ensuremath{\forall\;\Varid{a}\mathbin{.}Term_{FS}\;\Varid{f}\;\Varid{a}}, and the constructors of \ensuremath{Term_{FS}} are
hidden. Parametricity rules out \ensuremath{\Varid{badPlace}} and \ensuremath{\Varid{badCata}}, while the
use of an abstract type rules out \ensuremath{\Varid{badCase}}.

\subsection{Parametric Higher-Order Abstract Syntax}
\label{sec:phoas}

While the approach of Washburn and Weirich effectively rules out
exotic terms in Haskell, we prefer a different encoding that relies on
type parametricity only, and not an abstract type of terms. Our
solution is inspired by Chlipala's \emph{parametric higher-order
  abstract syntax} (PHOAS)~\cite{chlipala08icfp}. PHOAS is similar to
the restricted form of HOAS that we saw above; however, Chlipala makes
the parametricity explicit in the definition of terms by
distinguishing between the type of bound variables and the type of
recursive terms. In Chlipala's approach, an algebraic data type
encoding of lambda terms \ensuremath{\Conid{LTerm}} can effectively be defined via an
auxiliary data type \ensuremath{\Conid{LTrm}} of ``preterms'' as follows:
\begin{hscode}\SaveRestoreHook
\column{B}{@{}>{\hspre}l<{\hspost}@{}}%
\column{14}{@{}>{\hspre}c<{\hspost}@{}}%
\column{14E}{@{}l@{}}%
\column{17}{@{}>{\hspre}l<{\hspost}@{}}%
\column{E}{@{}>{\hspre}l<{\hspost}@{}}%
\>[B]{}\mathbf{type}\;\Conid{LTerm}{}\<[14]%
\>[14]{}\mathrel{=}{}\<[14E]%
\>[17]{}\forall\;\Varid{a}\mathbin{.}\Conid{LTrm}\;\Varid{a}{}\<[E]%
\\[\blanklineskip]%
\>[B]{}\mathbf{data}\;\Conid{LTrm}\;\Varid{a}{}\<[14]%
\>[14]{}\mathrel{=}{}\<[14E]%
\>[17]{}\Conid{Lam}\;(\Varid{a}\to \Conid{LTrm}\;\Varid{a})\mid \Conid{Var}\;\Varid{a}\mid \Conid{App}\;(\Conid{LTrm}\;\Varid{a})\;(\Conid{LTrm}\;\Varid{a}){}\<[E]%
\ColumnHook
\end{hscode}\resethooks
The definition of \ensuremath{\Conid{LTerm}} guarantees that all functions embedded via
\ensuremath{\Conid{Lam}} are parametric, and likewise that \ensuremath{\Conid{Var}}---Fegaras and Sheard's
\ensuremath{\Conid{Place}}---can only be applied to variables bound by an embedded
function. Atkey~\cite{atkey09tlca} showed that the encoding above
adequately captures closed lambda terms modulo $\alpha$-equivalence,
assuming that there is no infinite data and that all embedded
functions are total.

\subsubsection{Parametric Terms}
\label{sec:parametric-terms}

In order to transfer Chlipala's idea to non-recursive signatures and
catamorphisms, we need to distinguish between covariant and
contravariant uses of the recursive parameter. But this is exactly
what difunctors do! We therefore arrive at the following
definition of terms over difunctors:
\begin{hscode}\SaveRestoreHook
\column{B}{@{}>{\hspre}l<{\hspost}@{}}%
\column{17}{@{}>{\hspre}c<{\hspost}@{}}%
\column{17E}{@{}l@{}}%
\column{20}{@{}>{\hspre}l<{\hspost}@{}}%
\column{E}{@{}>{\hspre}l<{\hspost}@{}}%
\>[B]{}\mathbf{newtype}\;\Conid{Term}\;\Varid{f}{}\<[17]%
\>[17]{}\mathrel{=}{}\<[17E]%
\>[20]{}\Conid{Term}\;\{\mskip1.5mu \Varid{unTerm}\mathbin{::}\forall\;\Varid{a}\mathbin{.}\Conid{Trm}\;\Varid{f}\;\Varid{a}\mskip1.5mu\}{}\<[E]%
\\[\blanklineskip]%
\>[B]{}\mathbf{data}\;\Conid{Trm}\;\Varid{f}\;\Varid{a}{}\<[17]%
\>[17]{}\mathrel{=}{}\<[17E]%
\>[20]{}\Conid{In}\;(\Varid{f}\;\Varid{a}\;(\Conid{Trm}\;\Varid{f}\;\Varid{a}))\mid \Conid{Var}\;\Varid{a}\mbox{\onelinecomment  ``preterm''}{}\<[E]%
\ColumnHook
\end{hscode}\resethooks

Note the difference in \ensuremath{\Conid{Trm}} compared to \ensuremath{Term_{FS}} (besides using the
name \ensuremath{\Conid{Var}} rather than \ensuremath{\Conid{Place}}): the contravariant argument to the
difunctor \ensuremath{\Varid{f}} is not the type of terms \ensuremath{\Conid{Trm}\;\Varid{f}\;\Varid{a}}, but rather a
parametrised type \ensuremath{\Varid{a}}, which we quantify over at top-level to ensure
parametricity. Hence, the only way to use a bound variable is to wrap
it in a \ensuremath{\Conid{Var}} constructor---it is not possible to inspect the
parameter.  This representation more faithfully captures---we
believe---the restricted form of HOAS than the representation of
Washburn and Weirich: in our encoding it is explicit that bound
variables are merely placeholders, and not the same as
terms. Moreover, in some cases we actually \emph{need} to inspect the
structure of terms in order to define term transformations---we will
see such an example in Section~\ref{sec:term-transformations}. With an
abstract type of terms, this is not possible as Washburn and Weirich
note~\cite{washburn08jfp}.

Before we define algebras and catamorphisms, we lift the ideas
underlying CDTs to \emph{parametric compositional data types} (PCDTs),
namely coproducts and implicit injections. Fortunately, the
constructions of Section~\ref{sec:compositional-data-types} are
straightforwardly generalised (the instance declarations for \ensuremath{\fsub} are
exactly as in \dalc~\cite{swierstra08jfp}, so we omit them here):
\begin{hscode}\SaveRestoreHook
\column{B}{@{}>{\hspre}l<{\hspost}@{}}%
\column{3}{@{}>{\hspre}l<{\hspost}@{}}%
\column{9}{@{}>{\hspre}l<{\hspost}@{}}%
\column{19}{@{}>{\hspre}c<{\hspost}@{}}%
\column{19E}{@{}l@{}}%
\column{22}{@{}>{\hspre}l<{\hspost}@{}}%
\column{25}{@{}>{\hspre}l<{\hspost}@{}}%
\column{E}{@{}>{\hspre}l<{\hspost}@{}}%
\>[B]{}\mathbf{data}\;(\Varid{f}\fplus\Varid{g})\;\Varid{a}\;\Varid{b}\mathrel{=}\Conid{Inl}\;(\Varid{f}\;\Varid{a}\;\Varid{b})\mid \Conid{Inr}\;(\Varid{g}\;\Varid{a}\;\Varid{b}){}\<[E]%
\\[\blanklineskip]%
\>[B]{}\mathbf{instance}\;(\Conid{Difunctor}\;\Varid{f},\Conid{Difunctor}\;\Varid{g})\Rightarrow \Conid{Difunctor}\;(\Varid{f}\fplus\Varid{g})\;\mathbf{where}{}\<[E]%
\\
\>[B]{}\hsindent{3}{}\<[3]%
\>[3]{}\Varid{dimap}\;\Varid{f}\;\Varid{g}\;(\Conid{Inl}\;\Varid{x}){}\<[22]%
\>[22]{}\mathrel{=}{}\<[25]%
\>[25]{}\Conid{Inl}\;(\Varid{dimap}\;\Varid{f}\;\Varid{g}\;\Varid{x}){}\<[E]%
\\
\>[B]{}\hsindent{3}{}\<[3]%
\>[3]{}\Varid{dimap}\;\Varid{f}\;\Varid{g}\;(\Conid{Inr}\;\Varid{x}){}\<[22]%
\>[22]{}\mathrel{=}{}\<[25]%
\>[25]{}\Conid{Inr}\;(\Varid{dimap}\;\Varid{f}\;\Varid{g}\;\Varid{x}){}\<[E]%
\\[\blanklineskip]%
\>[B]{}\mathbf{class}\;\Varid{sub}\fsub\Varid{sup}\;\mathbf{where}{}\<[E]%
\\
\>[B]{}\hsindent{3}{}\<[3]%
\>[3]{}\Varid{inj}{}\<[9]%
\>[9]{}\mathbin{::}\Varid{sub}\;\Varid{a}\;\Varid{b}\to \Varid{sup}\;\Varid{a}\;\Varid{b}{}\<[E]%
\\
\>[B]{}\hsindent{3}{}\<[3]%
\>[3]{}\Varid{proj}{}\<[9]%
\>[9]{}\mathbin{::}\Varid{sup}\;\Varid{a}\;\Varid{b}\to \Conid{Maybe}\;(\Varid{sub}\;\Varid{a}\;\Varid{b}){}\<[E]%
\\[\blanklineskip]%
\>[B]{}\Varid{inject}\mathbin{::}(\Varid{g}\fsub\Varid{f})\Rightarrow \Varid{g}\;\Varid{a}\;(\Conid{Trm}\;\Varid{f}\;\Varid{a})\to \Conid{Trm}\;\Varid{f}\;\Varid{a}{}\<[E]%
\\
\>[B]{}\Varid{inject}\mathrel{=}\Conid{In}\mathbin{.}\Varid{inj}{}\<[E]%
\\[\blanklineskip]%
\>[B]{}\Varid{project}\mathbin{::}(\Varid{g}\fsub\Varid{f})\Rightarrow \Conid{Trm}\;\Varid{f}\;\Varid{a}\to \Conid{Maybe}\;(\Varid{g}\;\Varid{a}\;(\Conid{Trm}\;\Varid{f}\;\Varid{a})){}\<[E]%
\\
\>[B]{}\Varid{project}\;(\Conid{Term}\;\Varid{t}){}\<[19]%
\>[19]{}\mathrel{=}{}\<[19E]%
\>[22]{}\Varid{proj}\;\Varid{t}{}\<[E]%
\\
\>[B]{}\Varid{project}\;(\Conid{Var}\;\anonymous ){}\<[19]%
\>[19]{}\mathrel{=}{}\<[19E]%
\>[22]{}\Conid{Nothing}{}\<[E]%
\ColumnHook
\end{hscode}\resethooks
We can then recast our previous signatures from
Section~\ref{sec:motivating-example} as difunctors:
\begin{hscode}\SaveRestoreHook
\column{B}{@{}>{\hspre}l<{\hspost}@{}}%
\column{17}{@{}>{\hspre}c<{\hspost}@{}}%
\column{17E}{@{}l@{}}%
\column{20}{@{}>{\hspre}l<{\hspost}@{}}%
\column{49}{@{}>{\hspre}l<{\hspost}@{}}%
\column{64}{@{}>{\hspre}c<{\hspost}@{}}%
\column{64E}{@{}l@{}}%
\column{67}{@{}>{\hspre}l<{\hspost}@{}}%
\column{92}{@{}>{\hspre}l<{\hspost}@{}}%
\column{108}{@{}>{\hspre}c<{\hspost}@{}}%
\column{108E}{@{}l@{}}%
\column{111}{@{}>{\hspre}l<{\hspost}@{}}%
\column{E}{@{}>{\hspre}l<{\hspost}@{}}%
\>[B]{}\mathbf{data}\;\Conid{Lam}\;\Varid{a}\;\Varid{b}{}\<[17]%
\>[17]{}\mathrel{=}{}\<[17E]%
\>[20]{}\Conid{Lam}\;(\Varid{a}\to \Varid{b})$\qquad$\;{}\<[49]%
\>[49]{}\mathbf{data}\;\Conid{Lit}\;\Varid{a}\;\Varid{b}{}\<[64]%
\>[64]{}\mathrel{=}{}\<[64E]%
\>[67]{}\Conid{Lit}\;\Conid{Int}\;{}\<[92]%
\>[92]{}\mathbf{data}\;\Conid{Let}\;\Varid{a}\;\Varid{b}{}\<[108]%
\>[108]{}\mathrel{=}{}\<[108E]%
\>[111]{}\Conid{Let}\;\Varid{b}\;(\Varid{a}\to \Varid{b}){}\<[E]%
\\[\blanklineskip]%
\>[B]{}\mathbf{data}\;\Conid{App}\;\Varid{a}\;\Varid{b}{}\<[17]%
\>[17]{}\mathrel{=}{}\<[17E]%
\>[20]{}\Conid{App}\;\Varid{b}\;\Varid{b}\;{}\<[49]%
\>[49]{}\mathbf{data}\;\Conid{Plus}\;\Varid{a}\;\Varid{b}{}\<[64]%
\>[64]{}\mathrel{=}{}\<[64E]%
\>[67]{}\Conid{Plus}\;\Varid{b}\;\Varid{b}$\qquad$\;{}\<[92]%
\>[92]{}\mathbf{data}\;\Conid{Err}\;\Varid{a}\;\Varid{b}{}\<[108]%
\>[108]{}\mathrel{=}{}\<[108E]%
\>[111]{}\Conid{Err}{}\<[E]%
\\[\blanklineskip]%
\>[B]{}\mathbf{type}\;\Conid{Sig}{}\<[17]%
\>[17]{}\mathrel{=}{}\<[17E]%
\>[20]{}\Conid{Lam}\fplus\Conid{App}\fplus\Conid{Lit}\fplus\Conid{Plus}\fplus\Conid{Err}\fplus\Conid{Let}{}\<[E]%
\\[\blanklineskip]%
\>[B]{}\mathbf{type}\;\Conid{Sig'}{}\<[17]%
\>[17]{}\mathrel{=}{}\<[17E]%
\>[20]{}\Conid{Lam}\fplus\Conid{App}\fplus\Conid{Lit}\fplus\Conid{Plus}\fplus\Conid{Err}{}\<[E]%
\ColumnHook
\end{hscode}\resethooks

Finally, we can automatically derive instance declarations for
\ensuremath{\Conid{Difunctor}} as well as smart constructor definitions that comprise the
injections as for CDTs~\cite{bahr11wgp}. However, in order to avoid
the explicit \ensuremath{\Conid{Var}} constructor, we insert \ensuremath{\Varid{dimap}\;\Conid{Var}\;\Varid{id}} into the
declarations, e.g.\
\begin{hscode}\SaveRestoreHook
\column{B}{@{}>{\hspre}l<{\hspost}@{}}%
\column{E}{@{}>{\hspre}l<{\hspost}@{}}%
\>[B]{}\Varid{iLam}\mathbin{::}(\Conid{Lam}\fsub\Varid{f})\Rightarrow (\Conid{Trm}\;\Varid{f}\;\Varid{a}\to \Conid{Trm}\;\Varid{f}\;\Varid{a})\to \Conid{Trm}\;\Varid{f}\;\Varid{a}{}\<[E]%
\\
\>[B]{}\Varid{iLam}\;\Varid{f}\mathrel{=}\Varid{inject}\;(\Varid{dimap}\;\Conid{Var}\;\Varid{id}\;(\Conid{Lam}\;\Varid{f}))\mbox{\onelinecomment  (\ensuremath{\mathrel{=}\Varid{inject}\;(\Conid{Lam}\;(\Varid{f}\mathbin{.}\Conid{Var}))})}{}\<[E]%
\ColumnHook
\end{hscode}\resethooks
Using \ensuremath{\Varid{iLam}} we then need to be aware, though, that even if it takes a
function \ensuremath{\Conid{Trm}\;\Varid{f}\;\Varid{a}\to \Conid{Trm}\;\Varid{f}\;\Varid{a}} as argument, the input to that function
will always be of the form \ensuremath{\Conid{Var}\;\Varid{x}} \emph{by construction}. We can now again
represent terms such as $\textbf{let } x = 2 \textbf{ in } (\lambda
y. y + x)\; 3$ compactly as follows:\label{example-expression}
\begin{hscode}\SaveRestoreHook
\column{B}{@{}>{\hspre}l<{\hspost}@{}}%
\column{E}{@{}>{\hspre}l<{\hspost}@{}}%
\>[B]{}\Varid{e}\mathbin{::}\Conid{Term}\;\Conid{Sig}{}\<[E]%
\\
\>[B]{}\Varid{e}\mathrel{=}\Conid{Term}\;(\Varid{iLet}\;(\Varid{iLit}\;\mathrm{2})\;(\lambda \Varid{x}\to (\Varid{iLam}\;(\lambda \Varid{y}\to \Varid{y}\mathbin{`\Varid{iPlus}`}\Varid{x})\mathbin{`\Varid{iApp}`}\Varid{iLit}\;\mathrm{3}))){}\<[E]%
\ColumnHook
\end{hscode}\resethooks

\subsubsection{Algebras and Catamorphisms}
\label{sec:parametric-algebras-and-terms}

Given the representation of terms as fixed points of difunctors, we
can now define algebras and catamorphisms:
\begin{hscode}\SaveRestoreHook
\column{B}{@{}>{\hspre}l<{\hspost}@{}}%
\column{3}{@{}>{\hspre}l<{\hspost}@{}}%
\column{10}{@{}>{\hspre}l<{\hspost}@{}}%
\column{23}{@{}>{\hspre}c<{\hspost}@{}}%
\column{23E}{@{}l@{}}%
\column{26}{@{}>{\hspre}l<{\hspost}@{}}%
\column{E}{@{}>{\hspre}l<{\hspost}@{}}%
\>[B]{}\mathbf{type}\;\Conid{Alg}\;\Varid{f}\;\Varid{a}\mathrel{=}\Varid{f}\;\Varid{a}\;\Varid{a}\to \Varid{a}{}\<[E]%
\\[\blanklineskip]%
\>[B]{}\Varid{cata}\mathbin{::}\Conid{Difunctor}\;\Varid{f}\Rightarrow \Conid{Alg}\;\Varid{f}\;\Varid{a}\to \Conid{Term}\;\Varid{f}\to \Varid{a}{}\<[E]%
\\
\>[B]{}\Varid{cata}\;\phi\;(\Conid{Term}\;\Varid{t})\mathrel{=}\Varid{cat}\;\Varid{t}{}\<[E]%
\\
\>[B]{}\hsindent{3}{}\<[3]%
\>[3]{}\mathbf{where}\;{}\<[10]%
\>[10]{}\Varid{cat}\;(\Conid{In}\;\Varid{t}){}\<[23]%
\>[23]{}\mathrel{=}{}\<[23E]%
\>[26]{}\phi\;(\Varid{fmap}\;\Varid{cat}\;\Varid{t})\mbox{\onelinecomment  recall: \ensuremath{\Varid{fmap}\mathrel{=}\Varid{dimap}\;\Varid{id}}}{}\<[E]%
\\
\>[10]{}\Varid{cat}\;(\Conid{Var}\;\Varid{x}){}\<[23]%
\>[23]{}\mathrel{=}{}\<[23E]%
\>[26]{}\Varid{x}{}\<[E]%
\ColumnHook
\end{hscode}\resethooks

The definition of \ensuremath{\Varid{cata}} above is essentially the same as
\ensuremath{cata_{FS}}. The only difference is that bound variables within terms are
already wrapped in a \ensuremath{\Conid{Var}} constructor. Therefore, the contravariant
argument to \ensuremath{\Varid{dimap}} is the identity function, and we consequently use
the derived function \ensuremath{\Varid{fmap}} instead.

With these definitions in place, we can now recast the modular pretty
printer from Section~\ref{sec:motivating-example} to the new difunctor
signatures. However, since we now use a higher-order encoding, we need
to generate variable names for printing. We therefore arrive at the
following definition (the example is adopted from Washburn and
Weirich~\cite{washburn08jfp}, but we use streams rather than lists to
represent the sequence of available variable names):
\begin{hscode}\SaveRestoreHook
\column{B}{@{}>{\hspre}l<{\hspost}@{}}%
\column{3}{@{}>{\hspre}l<{\hspost}@{}}%
\column{5}{@{}>{\hspre}l<{\hspost}@{}}%
\column{40}{@{}>{\hspre}l<{\hspost}@{}}%
\column{E}{@{}>{\hspre}l<{\hspost}@{}}%
\>[B]{}\mathbf{data}\;\Conid{Stream}\;\Varid{a}\mathrel{=}\Conid{Cons}\;\Varid{a}\;(\Conid{Stream}\;\Varid{a}){}\<[E]%
\\[\blanklineskip]%
\>[B]{}\mathbf{class}\;\Conid{Pretty}\;\Varid{f}\;\mathbf{where}{}\<[E]%
\\
\>[B]{}\hsindent{3}{}\<[3]%
\>[3]{}\phi_{\mathrm{Pretty}}\mathbin{::}\Conid{Alg}\;\Varid{f}\;(\Conid{Stream}\;\Conid{String}\to \Conid{String}){}\<[E]%
\\[\blanklineskip]%
\>[B]{}\mbox{\onelinecomment  instance declaration that lifts \ensuremath{\Conid{Pretty}} to coproducts omitted}{}\<[E]%
\\[\blanklineskip]%
\>[B]{}\Varid{pretty}\mathbin{::}(\Conid{Difunctor}\;\Varid{f},\Conid{Pretty}\;\Varid{f})\Rightarrow \Conid{Term}\;\Varid{f}\to \Conid{String}{}\<[E]%
\\
\>[B]{}\Varid{pretty}\;\Varid{t}\mathrel{=}\Varid{cata}\;\phi_{\mathrm{Pretty}}\;\Varid{t}\;(\Varid{names}\;\mathrm{1}){}\<[E]%
\\
\>[B]{}\hsindent{5}{}\<[5]%
\>[5]{}\mathbf{where}\;\Varid{names}\;\Varid{n}\mathrel{=}\Conid{Cons}\;(\text{\tt 'x'}\mathbin{:}\Varid{show}\;\Varid{n})\;(\Varid{names}\;(\Varid{n}\mathbin{+}\mathrm{1})){}\<[E]%
\\[\blanklineskip]%
\>[B]{}\mathbf{instance}\;\Conid{Pretty}\;\Conid{Lam}\;\mathbf{where}{}\<[E]%
\\
\>[B]{}\hsindent{3}{}\<[3]%
\>[3]{}\phi_{\mathrm{Pretty}}\;(\Conid{Lam}\;\Varid{f})\;(\Conid{Cons}\;\Varid{x}\;\Varid{xs})\mathrel{=}\text{\tt \char34 (\char92 \char92 \char34}\plus \Varid{x}\plus \text{\tt \char34 .~\char34}\plus \Varid{f}\;(\Varid{const}\;\Varid{x})\;\Varid{xs}\plus \text{\tt \char34 )\char34}{}\<[E]%
\\[\blanklineskip]%
\>[B]{}\mathbf{instance}\;\Conid{Pretty}\;\Conid{App}\;\mathbf{where}{}\<[E]%
\\
\>[B]{}\hsindent{3}{}\<[3]%
\>[3]{}\phi_{\mathrm{Pretty}}\;(\Conid{App}\;\Varid{e}_{1}\;\Varid{e}_{2})\;\Varid{xs}\mathrel{=}\text{\tt \char34 (\char34}\plus \Varid{e}_{1}\;\Varid{xs}\plus \text{\tt \char34 ~\char34}\plus \Varid{e}_{2}\;\Varid{xs}\plus \text{\tt \char34 )\char34}{}\<[E]%
\\[\blanklineskip]%
\>[B]{}\mathbf{instance}\;\Conid{Pretty}\;\Conid{Lit}\;\mathbf{where}{}\<[E]%
\\
\>[B]{}\hsindent{3}{}\<[3]%
\>[3]{}\phi_{\mathrm{Pretty}}\;(\Conid{Lit}\;\Varid{n})\;\anonymous \mathrel{=}\Varid{show}\;\Varid{n}{}\<[E]%
\\[\blanklineskip]%
\>[B]{}\mathbf{instance}\;\Conid{Pretty}\;\Conid{Plus}\;\mathbf{where}{}\<[E]%
\\
\>[B]{}\hsindent{3}{}\<[3]%
\>[3]{}\phi_{\mathrm{Pretty}}\;(\Conid{Plus}\;\Varid{e}_{1}\;\Varid{e}_{2})\;\Varid{xs}\mathrel{=}\text{\tt \char34 (\char34}\plus \Varid{e}_{1}\;\Varid{xs}\plus \text{\tt \char34 ~+~\char34}\plus \Varid{e}_{2}\;\Varid{xs}\plus \text{\tt \char34 )\char34}{}\<[E]%
\\[\blanklineskip]%
\>[B]{}\mathbf{instance}\;\Conid{Pretty}\;\Conid{Let}\;\mathbf{where}{}\<[E]%
\\
\>[B]{}\hsindent{3}{}\<[3]%
\>[3]{}\phi_{\mathrm{Pretty}}\;(\Conid{Let}\;\Varid{e}_{1}\;\Varid{e}_{2})\;(\Conid{Cons}\;\Varid{x}\;\Varid{xs})\mathrel{=}{}\<[40]%
\>[40]{}\text{\tt \char34 (let~\char34}\plus \Varid{x}\plus \text{\tt \char34 ~=~\char34}\plus \Varid{e}_{1}\;\Varid{xs}\plus {}\<[E]%
\\
\>[40]{}\text{\tt \char34 ~in~\char34}\plus \Varid{e}_{2}\;(\Varid{const}\;\Varid{x})\;\Varid{xs}\plus \text{\tt \char34 )\char34}{}\<[E]%
\\[\blanklineskip]%
\>[B]{}\mathbf{instance}\;\Conid{Pretty}\;\Conid{Err}\;\mathbf{where}{}\<[E]%
\\
\>[B]{}\hsindent{3}{}\<[3]%
\>[3]{}\phi_{\mathrm{Pretty}}\;\Conid{Err}\;\anonymous \mathrel{=}\text{\tt \char34 error\char34}{}\<[E]%
\ColumnHook
\end{hscode}\resethooks
With this implementation of \ensuremath{\Varid{pretty}} we then have that \ensuremath{\Varid{pretty}\;\Varid{e}}
evaluates to the string \texttt{(let x1 = 2 in ((\char`\\ x2.\;(x2 +
  x1)) 3))}.

\subsubsection{Term Transformations}
\label{sec:term-transformations}
The pretty printer is an example of a modular computation over a
PCDT. However, we also want to define computations over PCDTs that
\emph{construct} PCDTs, e.g.\ the desugaring transformation. That is,
we want to construct functions of type \ensuremath{\Conid{Term}\;\Varid{f}\to \Conid{Term}\;\Varid{g}}, which means
that we must construct functions of type \ensuremath{(\forall\;\Varid{a}\mathbin{.}\Conid{Trm}\;\Varid{f}\;\Varid{a})\to (\forall\;\Varid{a}\mathbin{.}\Conid{Trm}\;\Varid{g}\;\Varid{a})}. Following the approach of
Section~\ref{sec:parametric-algebras-and-terms}, we construct such
functions by forming the catamorphisms of algebras of type \ensuremath{\Conid{Alg}\;\Varid{f}\;(\forall\;\Varid{a}\mathbin{.}\Conid{Trm}\;\Varid{g}\;\Varid{a})}, i.e.\ functions of type \ensuremath{\Varid{f}\;(\forall\;\Varid{a}\mathbin{.}\Conid{Trm}\;\Varid{g}\;\Varid{a})\;(\forall\;\Varid{a}\mathbin{.}\Conid{Trm}\;\Varid{g}\;\Varid{a})\to \forall\;\Varid{a}\mathbin{.}\Conid{Trm}\;\Varid{g}\;\Varid{a}}. However, in order to avoid the
nested quantifiers, we instead use \emph{parametric term algebras} of
type \ensuremath{\forall\;\Varid{a}\mathbin{.}\Conid{Alg}\;\Varid{f}\;(\Conid{Trm}\;\Varid{g}\;\Varid{a})}. From such algebras we then obtain
functions of the type \ensuremath{\forall\;\Varid{a}\mathbin{.}(\Conid{Trm}\;\Varid{f}\;\Varid{a}\to \Conid{Trm}\;\Varid{g}\;\Varid{a})} as
catamorphisms, which finally yield the desired functions of type \ensuremath{(\forall\;\Varid{a}\mathbin{.}\Conid{Trm}\;\Varid{f}\;\Varid{a})\to (\forall\;\Varid{a}\mathbin{.}\Conid{Trm}\;\Varid{g}\;\Varid{a})}. With these considerations in
mind, we arrive at the following definition of the desugaring algebra
type class:
\begin{hscode}\SaveRestoreHook
\column{B}{@{}>{\hspre}l<{\hspost}@{}}%
\column{3}{@{}>{\hspre}l<{\hspost}@{}}%
\column{E}{@{}>{\hspre}l<{\hspost}@{}}%
\>[B]{}\mathbf{class}\;\Conid{Desug}\;\Varid{f}\;\Varid{g}\;\mathbf{where}{}\<[E]%
\\
\>[B]{}\hsindent{3}{}\<[3]%
\>[3]{}\phi_{\mathrm{Desug}}\mathbin{::}\forall\;\Varid{a}\mathbin{.}\Conid{Alg}\;\Varid{f}\;(\Conid{Trm}\;\Varid{g}\;\Varid{a})\mbox{\onelinecomment  not \ensuremath{\Conid{Alg}\;\Varid{f}\;(\Conid{Term}\;\Varid{g})} !}{}\<[E]%
\\[\blanklineskip]%
\>[B]{}\mbox{\onelinecomment  instance declaration that lifts \ensuremath{\Conid{Desug}} to coproducts omitted}{}\<[E]%
\\[\blanklineskip]%
\>[B]{}\Varid{desug}\mathbin{::}(\Conid{Difunctor}\;\Varid{f},\Conid{Desug}\;\Varid{f}\;\Varid{g})\Rightarrow \Conid{Term}\;\Varid{f}\to \Conid{Term}\;\Varid{g}{}\<[E]%
\\
\>[B]{}\Varid{desug}\;\Varid{t}\mathrel{=}\Conid{Term}\;(\Varid{cata}\;\phi_{\mathrm{Desug}}\;\Varid{t}){}\<[E]%
\ColumnHook
\end{hscode}\resethooks
The algebra type class above is a \emph{multi-parameter type class}:
it is parametrised both by the domain signature \ensuremath{\Varid{f}} and the codomain
signature \ensuremath{\Varid{g}}. We do this in order to obtain a desugaring function
that is also modular in the codomain, similar to the evaluation
function for vanilla CDTs~\cite{bahr11wgp}.

We can now define the instances of \ensuremath{\Conid{Desug}} for the six signatures in
order to obtain the desugaring function. However, by utilising
overlapping instances we can make do with just two instance
declarations:
\begin{hscode}\SaveRestoreHook
\column{B}{@{}>{\hspre}l<{\hspost}@{}}%
\column{3}{@{}>{\hspre}l<{\hspost}@{}}%
\column{E}{@{}>{\hspre}l<{\hspost}@{}}%
\>[B]{}\mathbf{instance}\;(\Conid{Difunctor}\;\Varid{f},\Varid{f}\fsub\Varid{g})\Rightarrow \Conid{Desug}\;\Varid{f}\;\Varid{g}\;\mathbf{where}{}\<[E]%
\\
\>[B]{}\hsindent{3}{}\<[3]%
\>[3]{}\phi_{\mathrm{Desug}}\mathrel{=}\Varid{inject}\mathbin{.}\Varid{dimap}\;\Conid{Var}\;\Varid{id}\mbox{\onelinecomment  default instance for core signatures}{}\<[E]%
\\[\blanklineskip]%
\>[B]{}\mathbf{instance}\;(\Conid{App}\fsub\Varid{f},\Conid{Lam}\fsub\Varid{f})\Rightarrow \Conid{Desug}\;\Conid{Let}\;\Varid{f}\;\mathbf{where}{}\<[E]%
\\
\>[B]{}\hsindent{3}{}\<[3]%
\>[3]{}\phi_{\mathrm{Desug}}\;(\Conid{Let}\;\Varid{e}_{1}\;\Varid{e}_{2})\mathrel{=}\Varid{iLam}\;\Varid{e}_{2}\mathbin{`\Varid{iApp}`}\Varid{e}_{1}{}\<[E]%
\ColumnHook
\end{hscode}\resethooks
Given a term \ensuremath{\Varid{e}\mathbin{::}\Conid{Term}\;\Conid{Sig}}, we then have that \ensuremath{\Varid{desug}\;\Varid{e}\mathbin{::}\Conid{Term}\;\Conid{Sig'}}, i.e.\ the type shows that indeed all syntactic sugar has been
removed.

Whereas the desugaring transformation shows that we can construct
PCDTs from PCDTs in a modular fashion, we did not make use of the fact
that PCDTs can be inspected. That is, the desugaring transformation
does not inspect the recursively computed values, cf.\ the instance
declaration for \ensuremath{\Conid{Let}}. However, in order to implement the constant
folding transformation, we actually need to inspect recursively
computed PCDTs. We again utilise overlapping instances:
\begin{hscode}\SaveRestoreHook
\column{B}{@{}>{\hspre}l<{\hspost}@{}}%
\column{3}{@{}>{\hspre}l<{\hspost}@{}}%
\column{29}{@{}>{\hspre}l<{\hspost}@{}}%
\column{72}{@{}>{\hspre}l<{\hspost}@{}}%
\column{E}{@{}>{\hspre}l<{\hspost}@{}}%
\>[B]{}\mathbf{class}\;\Conid{Constf}\;\Varid{f}\;\Varid{g}\;\mathbf{where}{}\<[E]%
\\
\>[B]{}\hsindent{3}{}\<[3]%
\>[3]{}\phi_{\mathrm{Constf}}\mathbin{::}\forall\;\Varid{a}\mathbin{.}\Conid{Alg}\;\Varid{f}\;(\Conid{Trm}\;\Varid{g}\;\Varid{a}){}\<[E]%
\\[\blanklineskip]%
\>[B]{}\mbox{\onelinecomment  instance declaration that lifts \ensuremath{\Conid{Constf}} to coproducts omitted}{}\<[E]%
\\[\blanklineskip]%
\>[B]{}\Varid{constf}\mathbin{::}(\Conid{Difunctor}\;\Varid{f},\Conid{Constf}\;\Varid{f}\;\Varid{g})\Rightarrow \Conid{Term}\;\Varid{f}\to \Conid{Term}\;\Varid{g}{}\<[E]%
\\
\>[B]{}\Varid{constf}\;\Varid{t}\mathrel{=}\Conid{Term}\;(\Varid{cata}\;\phi_{\mathrm{Constf}}\;\Varid{t}){}\<[E]%
\\[\blanklineskip]%
\>[B]{}\mathbf{instance}\;(\Conid{Difunctor}\;\Varid{f},\Varid{f}\fsub\Varid{g})\Rightarrow \Conid{Constf}\;\Varid{f}\;\Varid{g}\;\mathbf{where}{}\<[E]%
\\
\>[B]{}\hsindent{3}{}\<[3]%
\>[3]{}\phi_{\mathrm{Constf}}\mathrel{=}\Varid{inject}\mathbin{.}\Varid{dimap}\;\Conid{Var}\;\Varid{id}\mbox{\onelinecomment  default instance}{}\<[E]%
\\[\blanklineskip]%
\>[B]{}\mathbf{instance}\;(\Conid{Plus}\fsub\Varid{f},\Conid{Lit}\fsub\Varid{f})\Rightarrow \Conid{Constf}\;\Conid{Plus}\;\Varid{f}\;\mathbf{where}{}\<[E]%
\\
\>[B]{}\hsindent{3}{}\<[3]%
\>[3]{}\phi_{\mathrm{Constf}}\;(\Conid{Plus}\;\Varid{e}_{1}\;\Varid{e}_{2})\mathrel{=}{}\<[29]%
\>[29]{}\mathbf{case}\;(\Varid{project}\;\Varid{e}_{1},\Varid{project}\;\Varid{e}_{2})\;\mathbf{of}{}\<[E]%
\\
\>[29]{}$\quad$(\Conid{Just}\;(\Conid{Lit}\;\Varid{n}),\Conid{Just}\;(\Conid{Lit}\;\Varid{m})){}\<[72]%
\>[72]{}\to \Varid{iLit}\;(\Varid{n}\mathbin{+}\Varid{m}){}\<[E]%
\\
\>[29]{}$\quad$\anonymous {}\<[72]%
\>[72]{}\to \Varid{e}_{1}\mathbin{`\Varid{iPlus}`}\Varid{e}_{2}{}\<[E]%
\ColumnHook
\end{hscode}\resethooks
Since we provide a default instance, we not
only obtain constant folding for the core language, but also for the
full language, i.e.\ \ensuremath{\Varid{constf}} has both the types \ensuremath{\Conid{Term}\;\Conid{Sig'}\to \Conid{Term}\;\Conid{Sig'}} and \ensuremath{\Conid{Term}\;\Conid{Sig}\to \Conid{Term}\;\Conid{Sig}}.

\section{Monadic Computations}
\label{sec:monadic-computations}

In the last section we demonstrated how to extend CDTs with
parametric higher-order abstract syntax, and how to perform modular,
recursive computations over terms containing binders. In this section
we investigate monadic computations over PCDTs.

\subsection{Monadic Interpretation}
\label{sec:monadic-interpretation}

While the previous examples of modular computations did not require
effects, the call-by-value interpreter prompts the need for monadic
computations: both in order to handle errors as well as 
controlling the evaluation order. Ultimately, we want to obtain a
function of the type \ensuremath{\Conid{Term}\;\Conid{Sig'}\to \Varid{m}\;(\Conid{Sem}\;\Varid{m})}, where the semantic
domain \ensuremath{\Conid{Sem}} is defined as follows (we use an ordinary algebraic data
type for simplicity):
\begin{hscode}\SaveRestoreHook
\column{B}{@{}>{\hspre}l<{\hspost}@{}}%
\column{E}{@{}>{\hspre}l<{\hspost}@{}}%
\>[B]{}\mathbf{data}\;\Conid{Sem}\;\Varid{m}\mathrel{=}\Conid{Fun}\;(\Conid{Sem}\;\Varid{m}\to \Varid{m}\;(\Conid{Sem}\;\Varid{m}))\mid \Conid{Int}\;\Conid{Int}{}\<[E]%
\ColumnHook
\end{hscode}\resethooks
Note that the monad only occurs in the codomain of \ensuremath{\Conid{Fun}}---if we want
call-by-name semantics rather than call-by-value semantics, we simply
add \ensuremath{\Varid{m}} also to the domain.

We can now implement the modular call-by-value interpreter similar to
the previous modular computations, but using the monadic algebra
carrier \ensuremath{\Varid{m}\;(\Conid{Sem}\;\Varid{m})}:
\begin{hscode}\SaveRestoreHook
\column{B}{@{}>{\hspre}l<{\hspost}@{}}%
\column{3}{@{}>{\hspre}l<{\hspost}@{}}%
\column{29}{@{}>{\hspre}l<{\hspost}@{}}%
\column{30}{@{}>{\hspre}l<{\hspost}@{}}%
\column{40}{@{}>{\hspre}l<{\hspost}@{}}%
\column{45}{@{}>{\hspre}l<{\hspost}@{}}%
\column{47}{@{}>{\hspre}l<{\hspost}@{}}%
\column{60}{@{}>{\hspre}c<{\hspost}@{}}%
\column{60E}{@{}l@{}}%
\column{64}{@{}>{\hspre}l<{\hspost}@{}}%
\column{E}{@{}>{\hspre}l<{\hspost}@{}}%
\>[B]{}\mathbf{class}\;\Conid{Monad}\;\Varid{m}\Rightarrow \Conid{Eval}\;\Varid{m}\;\Varid{f}\;\mathbf{where}{}\<[E]%
\\
\>[B]{}\hsindent{3}{}\<[3]%
\>[3]{}\phi_{\mathrm{Eval}}\mathbin{::}\Conid{Alg}\;\Varid{f}\;(\Varid{m}\;(\Conid{Sem}\;\Varid{m})){}\<[E]%
\\[\blanklineskip]%
\>[B]{}\mbox{\onelinecomment  instance declaration that lifts \ensuremath{\Conid{Eval}} to coproducts omitted}{}\<[E]%
\\[\blanklineskip]%
\>[B]{}\Varid{eval}\mathbin{::}(\Conid{Difunctor}\;\Varid{f},\Conid{Eval}\;\Varid{m}\;\Varid{f})\Rightarrow \Conid{Term}\;\Varid{f}\to \Varid{m}\;(\Conid{Sem}\;\Varid{m}){}\<[E]%
\\
\>[B]{}\Varid{eval}\mathrel{=}\Varid{cata}\;\phi_{\mathrm{Eval}}{}\<[E]%
\\[\blanklineskip]%
\>[B]{}\mathbf{instance}\;\Conid{Monad}\;\Varid{m}\Rightarrow \Conid{Eval}\;\Varid{m}\;\Conid{Lam}\;\mathbf{where}{}\<[E]%
\\
\>[B]{}\hsindent{3}{}\<[3]%
\>[3]{}\phi_{\mathrm{Eval}}\;(\Conid{Lam}\;\Varid{f})\mathrel{=}\Varid{return}\;(\Conid{Fun}\;(\Varid{f}\mathbin{.}\Varid{return})){}\<[E]%
\\[\blanklineskip]%
\>[B]{}\mathbf{instance}\;\Conid{MonadError}\;\Conid{String}\;\Varid{m}\Rightarrow \Conid{Eval}\;\Varid{m}\;\Conid{App}\;\mathbf{where}{}\<[E]%
\\
\>[B]{}\hsindent{3}{}\<[3]%
\>[3]{}\phi_{\mathrm{Eval}}\;(\Conid{App}\;\Varid{mx}\;\Varid{my})\mathrel{=}\mathbf{do}\;{}\<[29]%
\>[29]{}\Varid{x}\leftarrow \Varid{mx}{}\<[E]%
\\
\>[29]{}\mathbf{case}\;\Varid{x}\;\mathbf{of}\;{}\<[40]%
\>[40]{}\Conid{Fun}\;\Varid{f}{}\<[47]%
\>[47]{}\to \Varid{my}\bind \Varid{f}{}\<[E]%
\\
\>[40]{}\anonymous {}\<[47]%
\>[47]{}\to \Varid{throwError}\;\text{\tt \char34 stuck\char34}{}\<[E]%
\\[\blanklineskip]%
\>[B]{}\mathbf{instance}\;\Conid{Monad}\;\Varid{m}\Rightarrow \Conid{Eval}\;\Varid{m}\;\Conid{Lit}\;\mathbf{where}{}\<[E]%
\\
\>[B]{}\hsindent{3}{}\<[3]%
\>[3]{}\phi_{\mathrm{Eval}}\;(\Conid{Lit}\;\Varid{n})\mathrel{=}\Varid{return}\;(\Conid{Int}\;\Varid{n}){}\<[E]%
\\[\blanklineskip]%
\>[B]{}\mathbf{instance}\;\Conid{MonadError}\;\Conid{String}\;\Varid{m}\Rightarrow \Conid{Eval}\;\Varid{m}\;\Conid{Plus}\;\mathbf{where}{}\<[E]%
\\
\>[B]{}\hsindent{3}{}\<[3]%
\>[3]{}\phi_{\mathrm{Eval}}\;(\Conid{Plus}\;\Varid{mx}\;\Varid{my})\mathrel{=}\mathbf{do}\;{}\<[30]%
\>[30]{}\Varid{x}\leftarrow \Varid{mx}{}\<[E]%
\\
\>[30]{}\Varid{y}\leftarrow \Varid{my}{}\<[E]%
\\
\>[30]{}\mathbf{case}\;(\Varid{x},\Varid{y})\;\mathbf{of}\;{}\<[45]%
\>[45]{}(\Conid{Int}\;\Varid{n},\Conid{Int}\;\Varid{m}){}\<[60]%
\>[60]{}\to {}\<[60E]%
\>[64]{}\Varid{return}\;(\Conid{Int}\;(\Varid{n}\mathbin{+}\Varid{m})){}\<[E]%
\\
\>[45]{}\anonymous {}\<[60]%
\>[60]{}\to {}\<[60E]%
\>[64]{}\Varid{throwError}\;\text{\tt \char34 stuck\char34}{}\<[E]%
\\[\blanklineskip]%
\>[B]{}\mathbf{instance}\;\Conid{MonadError}\;\Conid{String}\;\Varid{m}\Rightarrow \Conid{Eval}\;\Varid{m}\;\Conid{Err}\;\mathbf{where}{}\<[E]%
\\
\>[B]{}\hsindent{3}{}\<[3]%
\>[3]{}\phi_{\mathrm{Eval}}\;\Conid{Err}\mathrel{=}\Varid{throwError}\;\text{\tt \char34 error\char34}{}\<[E]%
\ColumnHook
\end{hscode}\resethooks

In order to indicate errors in the course of the evaluation, we
require the monad to provide a method to throw an error. To this end,
we use the type class \ensuremath{\Conid{MonadError}}. Note how the modular design allows
us to require the stricter constraint \ensuremath{\Conid{MonadError}\;\Conid{String}\;\Varid{m}} only for
the cases where it is needed. This modularity of effects will become
quite useful when we will rule out \ensuremath{\text{\tt \char34 stuck\char34}} errors in
Section~\ref{sec:generalised-comp-data-types}.

With the interpreter definition above we have that \ensuremath{\Varid{eval}\;(\Varid{desug}\;\Varid{e})} evaluates to the value \ensuremath{\Conid{Right}\;(\Conid{Int}\;\mathrm{5})} as expected, where
\ensuremath{\Varid{e}} is as of page \pageref{example-expression} and \ensuremath{\Varid{m}} is the \ensuremath{\Conid{Either}\;\Conid{String}} monad. Moreover, we also have that $0 + \textbf{error}$ and $0
+ \lambda x. x$ evaluate to \ensuremath{\Conid{Left}\;\text{\tt \char34 error\char34}} and \ensuremath{\Conid{Left}\;\text{\tt \char34 stuck\char34}},
respectively.

\subsection{Monadic Computations with Implicit Sequencing}
\label{sec:monadic-comp-data}

In the example above we use a monadic algebra carrier for monadic
computations. For vanilla CDTs~\cite{bahr11wgp}, however, we have
previously shown how to perform monadic computations with
\emph{implicit sequencing}, by utilising the standard type class
\ensuremath{\Conid{Traversable}}\footnote{We have omitted methods from the definition of
  \ensuremath{\Conid{Traversable}} that are not necessary for our purposes.}:
\begin{hscode}\SaveRestoreHook
\column{B}{@{}>{\hspre}l<{\hspost}@{}}%
\column{3}{@{}>{\hspre}l<{\hspost}@{}}%
\column{36}{@{}>{\hspre}l<{\hspost}@{}}%
\column{E}{@{}>{\hspre}l<{\hspost}@{}}%
\>[B]{}\mathbf{type}\;\Conid{AlgM}\;\Varid{m}\;\Varid{f}\;\Varid{a}\mathrel{=}\Varid{f}\;\Varid{a}\to \Varid{m}\;\Varid{a}{}\<[E]%
\\[\blanklineskip]%
\>[B]{}\mathbf{class}\;\Conid{Functor}\;\Varid{f}\Rightarrow \Conid{Traversable}\;\Varid{f}\;\mathbf{where}{}\<[E]%
\\
\>[B]{}\hsindent{3}{}\<[3]%
\>[3]{}\Varid{sequence}\mathbin{::}\Conid{Monad}\;\Varid{m}\Rightarrow \Varid{f}\;(\Varid{m}\;\Varid{a})\to \Varid{m}\;(\Varid{f}\;\Varid{a}){}\<[E]%
\\[\blanklineskip]%
\>[B]{}\Varid{cataM}\mathbin{::}(\Conid{Traversable}\;\Varid{f},\Conid{Monad}\;\Varid{m}){}\<[36]%
\>[36]{}\Rightarrow \Conid{AlgM}\;\Varid{m}\;\Varid{f}\;\Varid{a}\to \Conid{Term}\;\Varid{f}\to \Varid{m}\;\Varid{a}{}\<[E]%
\\
\>[B]{}\Varid{cataM}\;\phi\mathrel{=}\phi\mathbin{<\!\!=\!\!<}\Varid{sequence}\mathbin{.}\Varid{fmap}\;(\Varid{cataM}\;\phi)\mathbin{.}\Varid{out}{}\<[E]%
\ColumnHook
\end{hscode}\resethooks

\ensuremath{\Conid{AlgM}\;\Varid{m}\;\Varid{f}\;\Varid{a}} represents the type of monadic
algebras~\cite{fokkinga94tr} over \ensuremath{\Varid{f}} and \ensuremath{\Varid{m}}, with carrier \ensuremath{\Varid{a}}, which
is different from \ensuremath{\Conid{Alg}\;\Varid{f}\;(\Varid{m}\;\Varid{a})} since the monad only occurs in the
codomain of the monadic algebra.  \ensuremath{\Varid{cataM}} is obtained from \ensuremath{\Varid{cata}} in
Section~\ref{sec:compositional-data-types} by performing \ensuremath{\Varid{sequence}}
after applying \ensuremath{\Varid{fmap}} and replacing function composition with monadic
function composition \ensuremath{\mathbin{<\!\!=\!\!<}}. That is, the recursion scheme takes care
of sequencing the monadic subcomputations.
Monadic algebras are useful for instance if we want to recursively
project a term over a compound signature to a smaller signature:
\begin{hscode}\SaveRestoreHook
\column{B}{@{}>{\hspre}l<{\hspost}@{}}%
\column{E}{@{}>{\hspre}l<{\hspost}@{}}%
\>[B]{}\Varid{deepProject}\mathbin{::}(\Conid{Traversable}\;\Varid{g},\Varid{f}\fsub\Varid{g})\Rightarrow \Conid{Term}\;\Varid{f}\to \Conid{Maybe}\;(\Conid{Term}\;\Varid{g}){}\<[E]%
\\
\>[B]{}\Varid{deepProject}\mathrel{=}\Varid{cataM}\;(\Varid{liftM}\;\Conid{In}\mathbin{.}\Varid{proj}){}\<[E]%
\ColumnHook
\end{hscode}\resethooks
Moreover, in a call-by-value setting we may use a monadic algebra \ensuremath{\Conid{Alg}\;\Varid{f}\;\Varid{m}\;\Varid{a}} rather than an ordinary algebra with a monadic carrier \ensuremath{\Conid{Alg}\;\Varid{f}\;(\Varid{m}\;\Varid{a})} in order to avoid the explicit sequencing of effects.

Turning back to parametric terms, we can apply the same idea to
difunctors yielding the following definition of monadic algebras:
\begin{hscode}\SaveRestoreHook
\column{B}{@{}>{\hspre}l<{\hspost}@{}}%
\column{E}{@{}>{\hspre}l<{\hspost}@{}}%
\>[B]{}\mathbf{type}\;\Conid{AlgM}\;\Varid{m}\;\Varid{f}\;\Varid{a}\mathrel{=}\Varid{f}\;\Varid{a}\;\Varid{a}\to \Varid{m}\;\Varid{a}{}\<[E]%
\ColumnHook
\end{hscode}\resethooks
Similarly, we can easily generalise \ensuremath{\Conid{Traversable}} and \ensuremath{\Varid{cataM}} to
difunctors:
\begin{hscode}\SaveRestoreHook
\column{B}{@{}>{\hspre}l<{\hspost}@{}}%
\column{3}{@{}>{\hspre}l<{\hspost}@{}}%
\column{26}{@{}>{\hspre}l<{\hspost}@{}}%
\column{29}{@{}>{\hspre}l<{\hspost}@{}}%
\column{36}{@{}>{\hspre}l<{\hspost}@{}}%
\column{38}{@{}>{\hspre}c<{\hspost}@{}}%
\column{38E}{@{}l@{}}%
\column{42}{@{}>{\hspre}l<{\hspost}@{}}%
\column{49}{@{}>{\hspre}l<{\hspost}@{}}%
\column{57}{@{}>{\hspre}l<{\hspost}@{}}%
\column{E}{@{}>{\hspre}l<{\hspost}@{}}%
\>[B]{}\mathbf{class}\;\Conid{Difunctor}\;\Varid{f}\Rightarrow \Conid{Ditraversable}\;\Varid{f}\;\mathbf{where}{}\<[E]%
\\
\>[B]{}\hsindent{3}{}\<[3]%
\>[3]{}\Varid{disequence}\mathbin{::}\Conid{Monad}\;\Varid{m}{}\<[26]%
\>[26]{}\Rightarrow \Varid{f}\;\Varid{a}\;(\Varid{m}\;\Varid{b})\to \Varid{m}\;(\Varid{f}\;\Varid{a}\;\Varid{b}){}\<[E]%
\\[\blanklineskip]%
\>[B]{}\Varid{cataM}\mathbin{::}(\Conid{Ditraversable}\;\Varid{f},\Conid{Monad}\;\Varid{m}){}\<[38]%
\>[38]{}\Rightarrow {}\<[38E]%
\>[42]{}\Conid{AlgM}\;\Varid{m}\;\Varid{f}\;\Varid{a}\to {}\<[57]%
\>[57]{}\Conid{Term}\;\Varid{f}\to \Varid{m}\;\Varid{a}{}\<[E]%
\\
\>[B]{}\Varid{cataM}\;\phi\;(\Conid{Term}\;\Varid{t})\mathrel{=}\Varid{cat}\;\Varid{t}\;{}\<[29]%
\>[29]{}\mathbf{where}\;{}\<[36]%
\>[36]{}\Varid{cat}\;(\Conid{In}\;\Varid{t}){}\<[49]%
\>[49]{}\mathrel{=}\Varid{disequence}\;(\Varid{fmap}\;\Varid{cat}\;\Varid{t})\bind \phi{}\<[E]%
\\
\>[36]{}\Varid{cat}\;(\Conid{Var}\;\Varid{x}){}\<[49]%
\>[49]{}\mathrel{=}\Varid{return}\;\Varid{x}{}\<[E]%
\ColumnHook
\end{hscode}\resethooks

Unfortunately, \ensuremath{\Varid{cataM}} only works for difunctors that do not use the
contravariant argument. To see why this is the case, reconsider the
\ensuremath{\Conid{Lam}} constructor; in order to define an instance of \ensuremath{\Conid{Ditraversable}}
for \ensuremath{\Conid{Lam}} we must write a function of the type:
\begin{hscode}\SaveRestoreHook
\column{B}{@{}>{\hspre}l<{\hspost}@{}}%
\column{E}{@{}>{\hspre}l<{\hspost}@{}}%
\>[B]{}\Varid{disequence}\mathbin{::}\Conid{Monad}\;\Varid{m}\Rightarrow \Conid{Lam}\;\Varid{a}\;(\Varid{m}\;\Varid{b})\to \Varid{m}\;(\Conid{Lam}\;\Varid{a}\;\Varid{b}){}\<[E]%
\ColumnHook
\end{hscode}\resethooks
Since \ensuremath{\Conid{Lam}} is isomorphic to the function type constructor \ensuremath{\to }, this
is equivalent to a function of the type:
\begin{hscode}\SaveRestoreHook
\column{B}{@{}>{\hspre}l<{\hspost}@{}}%
\column{E}{@{}>{\hspre}l<{\hspost}@{}}%
\>[B]{}\forall\;\Varid{a}\;\Varid{b}\;\Varid{m}\mathbin{.}\Conid{Monad}\;\Varid{m}\Rightarrow (\Varid{a}\to \Varid{m}\;\Varid{b})\to \Varid{m}\;(\Varid{a}\to \Varid{b}){}\<[E]%
\ColumnHook
\end{hscode}\resethooks
We cannot hope to be able to construct a meaningful combinator of that
type. Intuitively, in a function of type \ensuremath{\Varid{a}\to \Varid{m}\;\Varid{b}}, the monadic
effect of the result can depend on the input of type \ensuremath{\Varid{a}}. The monadic
effect of a monadic value of type \ensuremath{\Varid{m}\;(\Varid{a}\to \Varid{b})} is not dependent on
such input. For example, think of a state transformer monad \ensuremath{\Conid{ST}} with
state \ensuremath{\Conid{S}} and its put function \ensuremath{\Varid{put}\mathbin{::}\Conid{S}\to \Conid{ST}\;()}. What would be the
corresponding transformation to a monadic value of type \ensuremath{\Conid{ST}\;(\Conid{S}\to ())}?

Hence, \ensuremath{\Varid{cataM}} does not extend to terms with binders, but it still
works for terms without binders as in vanilla
CDTs~\cite{bahr11wgp}. In particular, we cannot use \ensuremath{\Varid{cataM}} to define
the call-by-value interpreter from
Section~\ref{sec:monadic-interpretation}.

\section{Contexts and Term Homomorphisms}
\label{sec:contexts-and-term-homomorphisms}

While the generality of catamorphisms makes them a powerful tool for
modular function definitions, their generality at the same time
inhibits flexibility and reusability. However, the full generality of
catamorphisms is not always needed in the case of term transformations,
which we discussed in Section~\ref{sec:term-transformations}. To this
end, we have previously studied term homomorphisms~\cite{bahr11wgp} as
a restricted form of term algebras.  In this section we redevelop
term homomorphisms for PCDTs.

\subsection{From Terms to Contexts and back}
\label{sec:contexts}

The crucial idea behind term homomorphisms is to generalise terms to
\emph{contexts}, i.e.\ terms with \emph{holes}. Following previous
work~\cite{bahr11wgp} we define the generalisation of terms with holes
as a \emph{generalised algebraic data type
  (GADT)}~\cite{schrijvers09icfp} with \emph{phantom types} \ensuremath{\Conid{Hole}} and
\ensuremath{\Conid{NoHole}}:
\begin{hscode}\SaveRestoreHook
\column{B}{@{}>{\hspre}l<{\hspost}@{}}%
\column{3}{@{}>{\hspre}l<{\hspost}@{}}%
\column{9}{@{}>{\hspre}l<{\hspost}@{}}%
\column{31}{@{}>{\hspre}l<{\hspost}@{}}%
\column{44}{@{}>{\hspre}l<{\hspost}@{}}%
\column{E}{@{}>{\hspre}l<{\hspost}@{}}%
\>[B]{}\mathbf{data}\;\Conid{Cxt}\mathbin{::}\mathbin{*}\to (\mathbin{*}\to \mathbin{*}\to \mathbin{*})\to \mathbin{*}\to \mathbin{*}\to \mathbin{*}\;\;\mathbf{where}{}\<[E]%
\\
\>[B]{}\hsindent{3}{}\<[3]%
\>[3]{}\Conid{In}{}\<[9]%
\>[9]{}\mathbin{::}\Varid{f}\;\Varid{a}\;(\Conid{Cxt}\;\Varid{h}\;\Varid{f}\;\Varid{a}\;\Varid{b}){}\<[31]%
\>[31]{}\to \Conid{Cxt}\;\Varid{h}\;{}\<[44]%
\>[44]{}\Varid{f}\;\Varid{a}\;\Varid{b}{}\<[E]%
\\
\>[B]{}\hsindent{3}{}\<[3]%
\>[3]{}\Conid{Var}{}\<[9]%
\>[9]{}\mathbin{::}\Varid{a}{}\<[31]%
\>[31]{}\to \Conid{Cxt}\;\Varid{h}\;{}\<[44]%
\>[44]{}\Varid{f}\;\Varid{a}\;\Varid{b}{}\<[E]%
\\
\>[B]{}\hsindent{3}{}\<[3]%
\>[3]{}\Conid{Hole}{}\<[9]%
\>[9]{}\mathbin{::}\Varid{b}{}\<[31]%
\>[31]{}\to \Conid{Cxt}\;\Conid{Hole}\;{}\<[44]%
\>[44]{}\Varid{f}\;\Varid{a}\;\Varid{b}{}\<[E]%
\\[\blanklineskip]%
\>[B]{}\mathbf{data}\;\Conid{Hole}{}\<[E]%
\\
\>[B]{}\mathbf{data}\;\Conid{NoHole}{}\<[E]%
\ColumnHook
\end{hscode}\resethooks

The first argument to \ensuremath{\Conid{Cxt}} is a phantom type indicating whether
the term contains holes or not. A context can thus be defined as:
\begin{hscode}\SaveRestoreHook
\column{B}{@{}>{\hspre}l<{\hspost}@{}}%
\column{E}{@{}>{\hspre}l<{\hspost}@{}}%
\>[B]{}\mathbf{type}\;\Conid{Context}\mathrel{=}\Conid{Cxt}\;\Conid{Hole}{}\<[E]%
\ColumnHook
\end{hscode}\resethooks
That is, contexts \emph{may} contain holes. On the other hand, terms
must not contain holes, so we can recover our previous definition of
preterms \ensuremath{\Conid{Trm}} as follows:
\begin{hscode}\SaveRestoreHook
\column{B}{@{}>{\hspre}l<{\hspost}@{}}%
\column{E}{@{}>{\hspre}l<{\hspost}@{}}%
\>[B]{}\mathbf{type}\;\Conid{Trm}\;\Varid{f}\;\Varid{a}\mathrel{=}\Conid{Cxt}\;\Conid{NoHole}\;\Varid{f}\;\Varid{a}\;(){}\<[E]%
\ColumnHook
\end{hscode}\resethooks
The definition of \ensuremath{\Conid{Term}} remains unchanged.
This representation of contexts and preterms allows us to uniformly
define functions that work on both types. For example, the function
\ensuremath{\Varid{inject}} now has the type:
\begin{hscode}\SaveRestoreHook
\column{B}{@{}>{\hspre}l<{\hspost}@{}}%
\column{E}{@{}>{\hspre}l<{\hspost}@{}}%
\>[B]{}\Varid{inject}\mathbin{::}(\Varid{g}\fsub\Varid{f})\Rightarrow \Varid{g}\;\Varid{a}\;(\Conid{Cxt}\;\Varid{h}\;\Varid{f}\;\Varid{a}\;\Varid{b})\to \Conid{Cxt}\;\Varid{h}\;\Varid{f}\;\Varid{a}\;\Varid{b}{}\<[E]%
\ColumnHook
\end{hscode}\resethooks

\subsection{Term Homomorphisms}
\label{sec:term-homomorphisms}

In Section~\ref{sec:term-transformations} we have shown that term
transformations, i.e.\ functions of type \ensuremath{\Conid{Term}\;\Varid{f}\to \Conid{Term}\;\Varid{g}}, are
obtained as catamorphisms of parametric term algebras of type \ensuremath{\forall\;\Varid{a}\mathbin{.}\Conid{Alg}\;\Varid{f}\;(\Conid{Trm}\;\Varid{g}\;\Varid{a})}. Spelling out the definition of \ensuremath{\Conid{Alg}}, such
algebras are functions of type:
\begin{hscode}\SaveRestoreHook
\column{B}{@{}>{\hspre}l<{\hspost}@{}}%
\column{E}{@{}>{\hspre}l<{\hspost}@{}}%
\>[B]{}\forall\;\Varid{a}\mathbin{.}\Varid{f}\;(\Conid{Trm}\;\Varid{g}\;\Varid{a})\;(\Conid{Trm}\;\Varid{g}\;\Varid{a})\to \Conid{Trm}\;\Varid{g}\;\Varid{a}{}\<[E]%
\ColumnHook
\end{hscode}\resethooks
As we have argued previously~\cite{bahr11wgp}, the fact that the
target signature \ensuremath{\Varid{g}} occurs in both the domain and codomain in the
above type prevents us from making use of the structure of the
algebra's carrier type \ensuremath{\Conid{Trm}\;\Varid{g}\;\Varid{a}}. In particular, the constructions
that we show in Section~\ref{sec:term-homom-resc} are not possible
with the above type.

In order to circumvent this restriction, we remove the occurrences of
the algebra's carrier type \ensuremath{\Conid{Trm}\;\Varid{g}\;\Varid{a}} in the domain by replacing them
with type variables:
\begin{hscode}\SaveRestoreHook
\column{B}{@{}>{\hspre}l<{\hspost}@{}}%
\column{E}{@{}>{\hspre}l<{\hspost}@{}}%
\>[B]{}\forall\;\Varid{a}\;\Varid{b}\mathbin{.}\Varid{f}\;\Varid{a}\;\Varid{b}\to \Conid{Trm}\;\Varid{g}\;\Varid{a}{}\<[E]%
\ColumnHook
\end{hscode}\resethooks
However, since we introduce a fresh variable \ensuremath{\Varid{b}}, functions of the
above type are not able to use the corresponding parts of the argument
for constructing the result. A value of type \ensuremath{\Varid{b}} cannot be injected
into the type \ensuremath{\Conid{Trm}\;\Varid{g}\;\Varid{a}}.

This is where contexts come into the picture: we enable the use of
values of type \ensuremath{\Varid{b}} in the result by replacing the codomain type \ensuremath{\Conid{Trm}\;\Varid{g}\;\Varid{a}} with \ensuremath{\Conid{Context}\;\Varid{g}\;\Varid{a}\;\Varid{b}}. The result is the following type of
\emph{term homomorphisms}:
\begin{hscode}\SaveRestoreHook
\column{B}{@{}>{\hspre}l<{\hspost}@{}}%
\column{E}{@{}>{\hspre}l<{\hspost}@{}}%
\>[B]{}\mathbf{type}\;\Conid{Hom}\;\Varid{f}\;\Varid{g}\mathrel{=}\forall\;\Varid{a}\;\Varid{b}\mathbin{.}\Varid{f}\;\Varid{a}\;\Varid{b}\to \Conid{Context}\;\Varid{g}\;\Varid{a}\;\Varid{b}{}\<[E]%
\ColumnHook
\end{hscode}\resethooks
A function \ensuremath{\rho\mathbin{::}\Conid{Hom}\;\Varid{f}\;\Varid{g}} is a transformation of constructors from
\ensuremath{\Varid{f}} into a context over \ensuremath{\Varid{g}}, i.e.\ a term over \ensuremath{\Varid{g}} that may embed
values taken from the arguments of the \ensuremath{\Varid{f}}-constructor. The parametric
polymorphism of the type guarantees that the arguments of the
\ensuremath{\Varid{f}}-constructor cannot be inspected but only embedded into the result
context. In order to apply term homomorphisms to terms, we need an
auxiliary function that merges nested contexts:
\begin{hscode}\SaveRestoreHook
\column{B}{@{}>{\hspre}l<{\hspost}@{}}%
\column{18}{@{}>{\hspre}l<{\hspost}@{}}%
\column{E}{@{}>{\hspre}l<{\hspost}@{}}%
\>[B]{}\Varid{appCxt}\mathbin{::}\Conid{Difunctor}\;\Varid{f}\Rightarrow \Conid{Context}\;\Varid{f}\;\Varid{a}\;(\Conid{Cxt}\;\Varid{h}\;\Varid{f}\;\Varid{a}\;\Varid{b})\to \Conid{Cxt}\;\Varid{h}\;\Varid{f}\;\Varid{a}\;\Varid{b}{}\<[E]%
\\
\>[B]{}\Varid{appCxt}\;(\Conid{In}\;\Varid{t}){}\<[18]%
\>[18]{}\mathrel{=}\Conid{In}\;(\Varid{fmap}\;\Varid{appCxt}\;\Varid{t}){}\<[E]%
\\
\>[B]{}\Varid{appCxt}\;(\Conid{Var}\;\Varid{x}){}\<[18]%
\>[18]{}\mathrel{=}\Conid{Var}\;\Varid{x}{}\<[E]%
\\
\>[B]{}\Varid{appCxt}\;(\Conid{Hole}\;\Varid{h}){}\<[18]%
\>[18]{}\mathrel{=}\Varid{h}{}\<[E]%
\ColumnHook
\end{hscode}\resethooks
Given a context that has terms embedded in its holes, we obtain a term
as a result; given a context with embedded contexts, the result is
again a context.

Using the combinator above we can now apply a term homomorphism to a
preterm---or more generally, to a context:
\begin{hscode}\SaveRestoreHook
\column{B}{@{}>{\hspre}l<{\hspost}@{}}%
\column{22}{@{}>{\hspre}l<{\hspost}@{}}%
\column{E}{@{}>{\hspre}l<{\hspost}@{}}%
\>[B]{}\Varid{appHom}\mathbin{::}(\Conid{Difunctor}\;\Varid{f},\Conid{Difunctor}\;\Varid{g})\Rightarrow \Conid{Hom}\;\Varid{f}\;\Varid{g}\to \Conid{Cxt}\;\Varid{h}\;\Varid{f}\;\Varid{a}\;\Varid{b}\to \Conid{Cxt}\;\Varid{h}\;\Varid{g}\;\Varid{a}\;\Varid{b}{}\<[E]%
\\
\>[B]{}\Varid{appHom}\;\rho\;(\Conid{In}\;\Varid{t}){}\<[22]%
\>[22]{}\mathrel{=}\Varid{appCxt}\;(\rho\;(\Varid{fmap}\;(\Varid{appHom}\;\rho)\;\Varid{t})){}\<[E]%
\\
\>[B]{}\Varid{appHom}\;\rho\;(\Conid{Var}\;\Varid{x}){}\<[22]%
\>[22]{}\mathrel{=}\Conid{Var}\;\Varid{x}{}\<[E]%
\\
\>[B]{}\Varid{appHom}\;\rho\;(\Conid{Hole}\;\Varid{h}){}\<[22]%
\>[22]{}\mathrel{=}\Conid{Hole}\;\Varid{h}{}\<[E]%
\ColumnHook
\end{hscode}\resethooks
From \ensuremath{\Varid{appHom}} we can then obtain the actual transformation on terms as
follows:
\begin{hscode}\SaveRestoreHook
\column{B}{@{}>{\hspre}l<{\hspost}@{}}%
\column{E}{@{}>{\hspre}l<{\hspost}@{}}%
\>[B]{}\Varid{appTHom}\mathbin{::}(\Conid{Difunctor}\;\Varid{f},\Conid{Difunctor}\;\Varid{g})\Rightarrow \Conid{Hom}\;\Varid{f}\;\Varid{g}\to \Conid{Term}\;\Varid{f}\to \Conid{Term}\;\Varid{g}{}\<[E]%
\\
\>[B]{}\Varid{appTHom}\;\rho\;(\Conid{Term}\;\Varid{t})\mathrel{=}\Conid{Term}\;(\Varid{appHom}\;\rho\;\Varid{t}){}\<[E]%
\ColumnHook
\end{hscode}\resethooks

Before we describe the benefits of term homomorphisms over term
algebras, we reconsider the desugaring transformation from
Section~\ref{sec:term-transformations}, but as a term homomorphism
rather than a term algebra:
\begin{hscode}\SaveRestoreHook
\column{B}{@{}>{\hspre}l<{\hspost}@{}}%
\column{3}{@{}>{\hspre}l<{\hspost}@{}}%
\column{E}{@{}>{\hspre}l<{\hspost}@{}}%
\>[B]{}\mathbf{class}\;\Conid{Desug}\;\Varid{f}\;\Varid{g}\;\mathbf{where}{}\<[E]%
\\
\>[B]{}\hsindent{3}{}\<[3]%
\>[3]{}\rho_{\mathrm{Desug}}\mathbin{::}\Conid{Hom}\;\Varid{f}\;\Varid{g}{}\<[E]%
\\[\blanklineskip]%
\>[B]{}\mbox{\onelinecomment  instance declaration that lifts \ensuremath{\Conid{Desug}} to coproducts omitted}{}\<[E]%
\\[\blanklineskip]%
\>[B]{}\Varid{desug}\mathbin{::}(\Conid{Difunctor}\;\Varid{f},\Conid{Difunctor}\;\Varid{g},\Conid{Desug}\;\Varid{f}\;\Varid{g})\Rightarrow \Conid{Term}\;\Varid{f}\to \Conid{Term}\;\Varid{g}{}\<[E]%
\\
\>[B]{}\Varid{desug}\mathrel{=}\Varid{appTHom}\;\rho_{\mathrm{Desug}}{}\<[E]%
\\[\blanklineskip]%
\>[B]{}\mathbf{instance}\;(\Conid{Difunctor}\;\Varid{f},\Conid{Difunctor}\;\Varid{g},\Varid{f}\fsub\Varid{g})\Rightarrow \Conid{Desug}\;\Varid{f}\;\Varid{g}\;\mathbf{where}{}\<[E]%
\\
\>[B]{}\hsindent{3}{}\<[3]%
\>[3]{}\rho_{\mathrm{Desug}}\mathrel{=}\Conid{In}\mathbin{.}\Varid{fmap}\;\Conid{Hole}\mathbin{.}\Varid{inj}\mbox{\onelinecomment  default instance for core signatures}{}\<[E]%
\\[\blanklineskip]%
\>[B]{}\mathbf{instance}\;(\Conid{App}\fsub\Varid{f},\Conid{Lam}\fsub\Varid{f})\Rightarrow \Conid{Desug}\;\Conid{Let}\;\Varid{f}\;\mathbf{where}{}\<[E]%
\\
\>[B]{}\hsindent{3}{}\<[3]%
\>[3]{}\rho_{\mathrm{Desug}}\;(\Conid{Let}\;\Varid{e}_{1}\;\Varid{e}_{2})\mathrel{=}\Varid{inject}\;(\Conid{Lam}\;(\Conid{Hole}\mathbin{.}\Varid{e}_{2}))\mathbin{`\Varid{iApp}`}\Conid{Hole}\;\Varid{e}_{1}{}\<[E]%
\ColumnHook
\end{hscode}\resethooks
Note how, in the instance declaration for \ensuremath{\Conid{Let}}, the constructor
\ensuremath{\Conid{Hole}} is used to embed arguments of the constructor \ensuremath{\Conid{Let}}, viz.\ \ensuremath{\Varid{e}_{1}}
and \ensuremath{\Varid{e}_{2}}, into the context that is constructed as the result.

As for the desugaring function in
Section~\ref{sec:term-transformations}, we utilise overlapping
instances to provide a default translation for the signatures that
need not be translated. The definitions above yield the desired
desugaring function \ensuremath{\Varid{desug}\mathbin{::}\Conid{Term}\;\Conid{Sig}\to \Conid{Term}\;\Conid{Sig'}}.

\subsection{Transforming and Combining Term Homomorphisms}
\label{sec:term-homom-resc}

In the following we shall shortly describe what we actually gain by
adopting the term homomorphism approach. First, term homomorphisms
enable automatic propagation of annotations, where annotations are
added via a restricted difunctor product, namely a product of a
difunctor \ensuremath{\Varid{f}} and a constant \ensuremath{\Varid{c}}:
\begin{hscode}\SaveRestoreHook
\column{B}{@{}>{\hspre}l<{\hspost}@{}}%
\column{E}{@{}>{\hspre}l<{\hspost}@{}}%
\>[B]{}\mathbf{data}\;(\Varid{f}\fann\Varid{c})\;\Varid{a}\;\Varid{b}\mathrel{=}\Varid{f}\;\Varid{a}\;\Varid{b}\fann\Varid{c}{}\<[E]%
\ColumnHook
\end{hscode}\resethooks
For instance, the type of ASTs of our language where each node is
annotated with source positions is captured by the type \ensuremath{\Conid{Term}\;(\Conid{Sig}\fann\Conid{SrcPos})}. With a term homomorphism \ensuremath{\Conid{Hom}\;\Varid{f}\;\Varid{g}} we automatically get a
lifted version \ensuremath{\Conid{Hom}\;(\Varid{f}\fann\Varid{c})\;(\Varid{g}\fann\Varid{c})}, which propagates annotations
from the input to the output. Hence, from our desugaring function in
the previous section we automatically get a lifted function on parse
trees \ensuremath{\Conid{Term}\;(\Conid{Sig}\fann\Conid{SrcPos})\to \Conid{Term}\;(\Conid{Sig'}\fann\Conid{SrcPos})},
which propagates source positions from the syntactic sugar
to the core constructs. We omit the details here, but note that the
constructions for CDTs~\cite{bahr11wgp} carry over straightforwardly
to PCDTs.

The second motivation for introducing term homomorphisms is
deforestation~\cite{wadler90tcs}. As we have shown
previously~\cite{bahr11wgp}, it is not possible to fuse two term
algebras in order to traverse the term only once. That is, we do not
find a composition operator \ensuremath{\circledcirc} on algebras that satisfies the
following equation:
\begin{center}
  \ensuremath{\Varid{cata}\;\phi_{1}\mathbin{.}\Varid{cata}\;\phi_{2}\mathrel{=}\Varid{cata}\;(\phi_{1}\circledcirc\phi_{2})} \quad for all \ensuremath{\phi_{1}\mathbin{::}\Conid{Alg}\;\Varid{g}\;\Varid{a}} and \ensuremath{\phi_{2}\mathbin{::}\forall\;\Varid{a}\mathbin{.}\Conid{Alg}\;\Varid{f}\;(\Conid{Trm}\;\Varid{g}\;\Varid{a})}
\end{center}
With term homomorphism, however, we do have such a composition
operator \ensuremath{\circledcirc}:
\begin{hscode}\SaveRestoreHook
\column{B}{@{}>{\hspre}l<{\hspost}@{}}%
\column{E}{@{}>{\hspre}l<{\hspost}@{}}%
\>[B]{}(\circledcirc)\mathbin{::}(\Conid{Difunctor}\;\Varid{g},\Conid{Difunctor}\;\Varid{h})\Rightarrow \Conid{Hom}\;\Varid{g}\;\Varid{h}\to \Conid{Hom}\;\Varid{f}\;\Varid{g}\to \Conid{Hom}\;\Varid{f}\;\Varid{h}{}\<[E]%
\\
\>[B]{}\rho_{1}\circledcirc\rho_{2}\mathrel{=}\Varid{appHom}\;\rho_{1}\mathbin{.}\rho_{2}{}\<[E]%
\ColumnHook
\end{hscode}\resethooks
For this composition, we then obtain the desired equation:
\begin{center}
  \ensuremath{\Varid{appHom}\;\rho_{1}\mathbin{.}\Varid{appHom}\;\rho_{2}\mathrel{=}\Varid{appHom}\;(\rho_{1}\circledcirc\rho_{2})} \quad for all \ensuremath{\rho_{1}\mathbin{::}\Conid{Hom}\;\Varid{g}\;\Varid{h}} and \ensuremath{\rho_{2}\mathbin{::}\Conid{Hom}\;\Varid{f}\;\Varid{g}}
\end{center}

In fact, we can also compose an arbitrary algebra with a term
homomorphism:
\begin{hscode}\SaveRestoreHook
\column{B}{@{}>{\hspre}l<{\hspost}@{}}%
\column{23}{@{}>{\hspre}c<{\hspost}@{}}%
\column{23E}{@{}l@{}}%
\column{27}{@{}>{\hspre}l<{\hspost}@{}}%
\column{50}{@{}>{\hspre}l<{\hspost}@{}}%
\column{E}{@{}>{\hspre}l<{\hspost}@{}}%
\>[B]{}(\boxdot)\mathbin{::}\Conid{Difunctor}\;\Varid{g}{}\<[23]%
\>[23]{}\Rightarrow {}\<[23E]%
\>[27]{}\Conid{Alg}\;\Varid{g}\;\Varid{a}\to \Conid{Hom}\;\Varid{f}\;\Varid{g}\to {}\<[50]%
\>[50]{}\Conid{Alg}\;\Varid{f}\;\Varid{a}{}\<[E]%
\\
\>[B]{}\phi\boxdot\rho\mathrel{=}\Varid{free}\;\phi\;\Varid{id}\mathbin{.}\rho{}\<[E]%
\ColumnHook
\end{hscode}\resethooks
where
\begin{hscode}\SaveRestoreHook
\column{B}{@{}>{\hspre}l<{\hspost}@{}}%
\column{11}{@{}>{\hspre}l<{\hspost}@{}}%
\column{14}{@{}>{\hspre}l<{\hspost}@{}}%
\column{24}{@{}>{\hspre}l<{\hspost}@{}}%
\column{E}{@{}>{\hspre}l<{\hspost}@{}}%
\>[B]{}\Varid{free}\mathbin{::}\Conid{Difunctor}\;\Varid{f}\Rightarrow \Conid{Alg}\;\Varid{f}\;\Varid{a}\to (\Varid{b}\to \Varid{a})\to \Conid{Cxt}\;\Varid{h}\;\Varid{f}\;\Varid{a}\;\Varid{b}\to \Varid{a}{}\<[E]%
\\
\>[B]{}\Varid{free}\;\phi\;{}\<[11]%
\>[11]{}\Varid{f}\;{}\<[14]%
\>[14]{}(\Conid{In}\;\Varid{t}){}\<[24]%
\>[24]{}\mathrel{=}\phi\;(\Varid{fmap}\;(\Varid{free}\;\phi\;\Varid{f})\;\Varid{t}){}\<[E]%
\\
\>[B]{}\Varid{free}\;\anonymous \;{}\<[11]%
\>[11]{}\anonymous \;{}\<[14]%
\>[14]{}(\Conid{Var}\;\Varid{x}){}\<[24]%
\>[24]{}\mathrel{=}\Varid{x}{}\<[E]%
\\
\>[B]{}\Varid{free}\;\anonymous \;{}\<[11]%
\>[11]{}\Varid{f}\;{}\<[14]%
\>[14]{}(\Conid{Hole}\;\Varid{h}){}\<[24]%
\>[24]{}\mathrel{=}\Varid{f}\;\Varid{h}{}\<[E]%
\ColumnHook
\end{hscode}\resethooks
The composition of algebras and homomorphisms satisfies the following equation:
\begin{center}
  \ensuremath{\Varid{cata}\;\phi\mathbin{.}\Varid{appHom}\;\rho\mathrel{=}\Varid{cata}\;(\phi\boxdot\rho)} \quad for all \ensuremath{\phi\mathbin{::}\Conid{Alg}\;\Varid{g}\;\Varid{a}} and \ensuremath{\rho\mathbin{::}\Conid{Hom}\;\Varid{f}\;\Varid{g}}
\end{center}

For example, in order to evaluate a term with syntactic sugar, rather
than composing \ensuremath{\Varid{eval}} and \ensuremath{\Varid{desug}}, we can use the function \ensuremath{\Varid{cata}\;(\phi_{\mathrm{Eval}}\boxdot\rho_{\mathrm{Desug}})}, which only traverses the term once. This
transformation can be automated using GHC's rewrite
mechanism~\cite{jones01haskell} and our experimental results for CDTs
show that the thus obtained speedup is significant~\cite{bahr11wgp}.

\section{Generalised Parametric Compositional Data Types}
\label{sec:generalised-comp-data-types}

In this section we briefly describe how to lift the construction of
mutually recursive data types and---more generally---GADTs from CDTs
to PCDTs. The construction is based on the work of Johann and
Ghani~\cite{johann08popl}. For CDTs the generalisation, roughly
speaking, amounts to lifting functors to (generalised)
\emph{higher-order functors}~\cite{johann08popl}, and functions on
terms to \emph{natural transformations}, as shown
earlier~\cite{bahr11wgp}:
\begin{hscode}\SaveRestoreHook
\column{B}{@{}>{\hspre}l<{\hspost}@{}}%
\column{3}{@{}>{\hspre}l<{\hspost}@{}}%
\column{E}{@{}>{\hspre}l<{\hspost}@{}}%
\>[B]{}\mathbf{type}\;\Varid{a}\nattrans\Varid{b}\mathrel{=}\forall\;\Varid{i}\mathbin{.}\Varid{a}\;\Varid{i}\to \Varid{b}\;\Varid{i}{}\<[E]%
\\[\blanklineskip]%
\>[B]{}\mathbf{class}\;\Conid{HFunctor}\;\Varid{f}\;\mathbf{where}{}\<[E]%
\\
\>[B]{}\hsindent{3}{}\<[3]%
\>[3]{}\Varid{hfmap}\mathbin{::}\Varid{a}\nattrans\Varid{b}\to \Varid{f}\;\Varid{a}\nattrans\Varid{f}\;\Varid{b}{}\<[E]%
\ColumnHook
\end{hscode}\resethooks
Now, signatures are of the kind \ensuremath{(\mathbin{*}\to \mathbin{*})\to \mathbin{*}\to \mathbin{*}}, rather than \ensuremath{\mathbin{*}\to \mathbin{*}}, which reflects the fact that signatures are now \emph{indexed
  types}, and so are terms (or contexts in general). Consequently, the
carrier of an algebra is a type constructor of kind \ensuremath{\mathbin{*}\to \mathbin{*}}:
\begin{hscode}\SaveRestoreHook
\column{B}{@{}>{\hspre}l<{\hspost}@{}}%
\column{E}{@{}>{\hspre}l<{\hspost}@{}}%
\>[B]{}\mathbf{type}\;\Conid{Alg}\;\Varid{f}\;\Varid{a}\mathrel{=}\Varid{f}\;\Varid{a}\nattrans\Varid{a}{}\<[E]%
\ColumnHook
\end{hscode}\resethooks
Since signatures will be defined as GADTs, we effectively deal with
\emph{many-sorted algebras}. If a subterm has the type index \ensuremath{\Varid{i}}, then
the value computed recursively by a catamorphism will have the type \ensuremath{\Varid{a}\;\Varid{i}}. The coproduct \ensuremath{\fplus} and the automatic injections \ensuremath{\fsub} carry over
straightforwardly from functors to higher-order
functors~\cite{bahr11wgp}.

In order to lift the ideas from CDTs to PCDTs, we need to consider
indexed difunctors. This prompts the notion of \emph{higher-order
  difunctors}:
\begin{hscode}\SaveRestoreHook
\column{B}{@{}>{\hspre}l<{\hspost}@{}}%
\column{3}{@{}>{\hspre}l<{\hspost}@{}}%
\column{E}{@{}>{\hspre}l<{\hspost}@{}}%
\>[B]{}\mathbf{class}\;\Conid{HDifunctor}\;\Varid{f}\;\mathbf{where}{}\<[E]%
\\
\>[B]{}\hsindent{3}{}\<[3]%
\>[3]{}\Varid{hdimap}\mathbin{::}(\Varid{a}\nattrans\Varid{b})\to (\Varid{c}\nattrans\Varid{d})\to \Varid{f}\;\Varid{b}\;\Varid{c}\nattrans\Varid{f}\;\Varid{a}\;\Varid{d}{}\<[E]%
\\[\blanklineskip]%
\>[B]{}\mathbf{instance}\;\Conid{HDifunctor}\;\Varid{f}\Rightarrow \Conid{HFunctor}\;(\Varid{f}\;\Varid{a})\;\mathbf{where}{}\<[E]%
\\
\>[B]{}\hsindent{3}{}\<[3]%
\>[3]{}\Varid{hfmap}\mathrel{=}\Varid{hdimap}\;\Varid{id}{}\<[E]%
\ColumnHook
\end{hscode}\resethooks
Note the familiar pattern from ordinary PCDTs: a higher-order
difunctor gives rise to a higher-order functor when the contravariant
argument is fixed.

To illustrate higher-order difunctors, consider a modular GADT
encoding of our core language:
\begin{hscode}\SaveRestoreHook
\column{B}{@{}>{\hspre}l<{\hspost}@{}}%
\column{3}{@{}>{\hspre}l<{\hspost}@{}}%
\column{E}{@{}>{\hspre}l<{\hspost}@{}}%
\>[B]{}\mathbf{data}\;\Conid{TArrow}\;\Varid{i}\;\Varid{j}{}\<[E]%
\\[\blanklineskip]%
\>[B]{}\mathbf{data}\;\Conid{TInt}{}\<[E]%
\\[\blanklineskip]%
\>[B]{}\mathbf{data}\;\Conid{Lam}\mathbin{::}(\mathbin{*}\to \mathbin{*})\to (\mathbin{*}\to \mathbin{*})\to \mathbin{*}\to \mathbin{*}\;\;\mathbf{where}{}\<[E]%
\\
\>[B]{}\hsindent{3}{}\<[3]%
\>[3]{}\Conid{Lam}\mathbin{::}(\Varid{a}\;\Varid{i}\to \Varid{b}\;\Varid{j})\to \Conid{Lam}\;\Varid{a}\;\Varid{b}\;(\Varid{i}\mathbin{`\Conid{TArrow}`}\Varid{j}){}\<[E]%
\\[\blanklineskip]%
\>[B]{}\mathbf{data}\;\Conid{App}\mathbin{::}(\mathbin{*}\to \mathbin{*})\to (\mathbin{*}\to \mathbin{*})\to \mathbin{*}\to \mathbin{*}\;\;\mathbf{where}{}\<[E]%
\\
\>[B]{}\hsindent{3}{}\<[3]%
\>[3]{}\Conid{App}\mathbin{::}\Varid{b}\;(\Varid{i}\mathbin{`\Conid{TArrow}`}\Varid{j})\to \Varid{b}\;\Varid{i}\to \Conid{App}\;\Varid{a}\;\Varid{b}\;\Varid{j}{}\<[E]%
\\[\blanklineskip]%
\>[B]{}\mathbf{data}\;\Conid{Lit}\mathbin{::}(\mathbin{*}\to \mathbin{*})\to (\mathbin{*}\to \mathbin{*})\to \mathbin{*}\to \mathbin{*}\;\;\mathbf{where}{}\<[E]%
\\
\>[B]{}\hsindent{3}{}\<[3]%
\>[3]{}\Conid{Lit}\mathbin{::}\Conid{Int}\to \Conid{Lit}\;\Varid{a}\;\Varid{b}\;\Conid{TInt}{}\<[E]%
\\[\blanklineskip]%
\>[B]{}\mathbf{data}\;\Conid{Plus}\mathbin{::}(\mathbin{*}\to \mathbin{*})\to (\mathbin{*}\to \mathbin{*})\to \mathbin{*}\to \mathbin{*}\;\;\mathbf{where}{}\<[E]%
\\
\>[B]{}\hsindent{3}{}\<[3]%
\>[3]{}\Conid{Plus}\mathbin{::}\Varid{b}\;\Conid{TInt}\to \Varid{b}\;\Conid{TInt}\to \Conid{Plus}\;\Varid{a}\;\Varid{b}\;\Conid{TInt}{}\<[E]%
\\[\blanklineskip]%
\>[B]{}\mathbf{data}\;\Conid{Err}\mathbin{::}(\mathbin{*}\to \mathbin{*})\to (\mathbin{*}\to \mathbin{*})\to \mathbin{*}\to \mathbin{*}\;\;\mathbf{where}{}\<[E]%
\\
\>[B]{}\hsindent{3}{}\<[3]%
\>[3]{}\Conid{Err}\mathbin{::}\Conid{Err}\;\Varid{a}\;\Varid{b}\;\Varid{i}{}\<[E]%
\\[\blanklineskip]%
\>[B]{}\mathbf{type}\;\Conid{Sig'}\mathrel{=}\Conid{Lam}\fplus\Conid{App}\fplus\Conid{Lit}\fplus\Conid{Plus}\fplus\Conid{Err}{}\<[E]%
\ColumnHook
\end{hscode}\resethooks
Note, in particular, the type of \ensuremath{\Conid{Lam}}: now the bound variable is
typed!

We use \ensuremath{\Conid{TArrow}} and \ensuremath{\Conid{TInt}} as type indices for the GADT definitions
above. The preference of these fresh types over Haskell's \ensuremath{\to } and
\ensuremath{\Conid{Int}} is meant to emphasise that these phantom types are only labels
that represent the type constructors of our object language.

We use the coproduct \ensuremath{\fplus} of higher-order difunctors above to combine
signatures, which is easily defined, and as for CDTs it is
straightforward to lift instances of \ensuremath{\Conid{HDifunctor}} for \ensuremath{\Varid{f}} and \ensuremath{\Varid{g}} to
an instance for \ensuremath{\Varid{f}\fplus\Varid{g}}. Similarly, we can generalise the relation
\ensuremath{\fsub} from difunctors to higher-order difunctors, so we omit its
definition here.

The type of generalised parametric (pre)terms can now be constructed
as an indexed type:
\begin{hscode}\SaveRestoreHook
\column{B}{@{}>{\hspre}l<{\hspost}@{}}%
\column{19}{@{}>{\hspre}l<{\hspost}@{}}%
\column{E}{@{}>{\hspre}l<{\hspost}@{}}%
\>[B]{}\mathbf{newtype}\;\Conid{Term}\;\Varid{f}\;\Varid{i}{}\<[19]%
\>[19]{}\mathrel{=}\Conid{Term}\;\{\mskip1.5mu \Varid{unTerm}\mathbin{::}\forall\;\Varid{a}\mathbin{.}\Conid{Trm}\;\Varid{f}\;\Varid{a}\;\Varid{i}\mskip1.5mu\}{}\<[E]%
\\
\>[B]{}\mathbf{data}\;\Conid{Trm}\;\Varid{f}\;\Varid{a}\;\Varid{i}{}\<[19]%
\>[19]{}\mathrel{=}\Conid{In}\;(\Varid{f}\;\Varid{a}\;(\Conid{Trm}\;\Varid{f}\;\Varid{a})\;\Varid{i})\mid \Conid{Var}\;(\Varid{a}\;\Varid{i}){}\<[E]%
\ColumnHook
\end{hscode}\resethooks
Moreover, we use smart constructors as for PCDTs to compactly
construct terms, for instance:
\begin{hscode}\SaveRestoreHook
\column{B}{@{}>{\hspre}l<{\hspost}@{}}%
\column{E}{@{}>{\hspre}l<{\hspost}@{}}%
\>[B]{}\Varid{e}\mathbin{::}\Conid{Term}\;\Conid{Sig'}\;\Conid{TInt}{}\<[E]%
\\
\>[B]{}\Varid{e}\mathrel{=}\Conid{Term}\;(\Varid{iLam}\;(\lambda \Varid{x}\to \Varid{x}\mathbin{`\Varid{iPlus}`}\Varid{x})\mathbin{`\Varid{iApp}`}\Varid{iLit}\;\mathrm{2}){}\<[E]%
\ColumnHook
\end{hscode}\resethooks

Finally, we can lift algebras and their induced catamorphisms by
lifting the definitions in
Section~\ref{sec:parametric-algebras-and-terms} via natural
transformations and higher-order difunctors:
\begin{hscode}\SaveRestoreHook
\column{B}{@{}>{\hspre}l<{\hspost}@{}}%
\column{5}{@{}>{\hspre}l<{\hspost}@{}}%
\column{12}{@{}>{\hspre}l<{\hspost}@{}}%
\column{25}{@{}>{\hspre}c<{\hspost}@{}}%
\column{25E}{@{}l@{}}%
\column{28}{@{}>{\hspre}l<{\hspost}@{}}%
\column{E}{@{}>{\hspre}l<{\hspost}@{}}%
\>[B]{}\mathbf{type}\;\Conid{Alg}\;\Varid{f}\;\Varid{a}\mathrel{=}\Varid{f}\;\Varid{a}\;\Varid{a}\nattrans\Varid{a}{}\<[E]%
\\[\blanklineskip]%
\>[B]{}\Varid{cata}\mathbin{::}\Conid{HDifunctor}\;\Varid{f}\Rightarrow \Conid{Alg}\;\Varid{f}\;\Varid{a}\to \Conid{Term}\;\Varid{f}\nattrans\Varid{a}{}\<[E]%
\\
\>[B]{}\Varid{cata}\;\phi\;(\Conid{Term}\;\Varid{t})\mathrel{=}\Varid{cat}\;\Varid{t}{}\<[E]%
\\
\>[B]{}\hsindent{5}{}\<[5]%
\>[5]{}\mathbf{where}\;{}\<[12]%
\>[12]{}\Varid{cat}\;(\Conid{In}\;\Varid{t}){}\<[25]%
\>[25]{}\mathrel{=}{}\<[25E]%
\>[28]{}\phi\;(\Varid{hfmap}\;\Varid{cat}\;\Varid{t})\mbox{\onelinecomment  recall: \ensuremath{\Varid{hfmap}\mathrel{=}\Varid{hdimap}\;\Varid{id}}}{}\<[E]%
\\
\>[12]{}\Varid{cat}\;(\Conid{Var}\;\Varid{x}){}\<[25]%
\>[25]{}\mathrel{=}{}\<[25E]%
\>[28]{}\Varid{x}{}\<[E]%
\ColumnHook
\end{hscode}\resethooks

With the definitions above we can now define a call-by-value
interpreter for our typed example language. To this end, we must
provide a type-level function that, for a given object language type
constructed from \ensuremath{\Conid{TArrow}} and \ensuremath{\Conid{TInt}}, selects the corresponding subset
of the semantic domain \ensuremath{\Conid{Sem}\;\Varid{m}} from
Section~\ref{sec:monadic-interpretation}. This can be achieved via
Haskell's \emph{type families}~\cite{schrijvers08icfp}:
\begin{hscode}\SaveRestoreHook
\column{B}{@{}>{\hspre}l<{\hspost}@{}}%
\column{37}{@{}>{\hspre}l<{\hspost}@{}}%
\column{E}{@{}>{\hspre}l<{\hspost}@{}}%
\>[B]{}\mathbf{type}\;\textbf{family}\;\Conid{Sem}\;(\Varid{m}\mathbin{::}\mathbin{*}\to \mathbin{*})\;\Varid{i}{}\<[E]%
\\
\>[B]{}\mathbf{type}\;\mathbf{instance}\;\Conid{Sem}\;\Varid{m}\;(\Varid{i}\mathbin{`\Conid{TArrow}`}\Varid{j}){}\<[37]%
\>[37]{}\mathrel{=}\Conid{Sem}\;\Varid{m}\;\Varid{i}\to \Varid{m}\;(\Conid{Sem}\;\Varid{m}\;\Varid{j}){}\<[E]%
\\
\>[B]{}\mathbf{type}\;\mathbf{instance}\;\Conid{Sem}\;\Varid{m}\;\Conid{TInt}{}\<[37]%
\>[37]{}\mathrel{=}\Conid{Int}{}\<[E]%
\ColumnHook
\end{hscode}\resethooks
The type \ensuremath{\Conid{Sem}\;\Varid{m}\;\Varid{t}} is obtained from an object language type \ensuremath{\Varid{t}} by
replacing each function type \ensuremath{\Varid{t}_{1}\mathbin{`\Conid{TArrow}`}\Varid{t}_{2}} occurring in \ensuremath{\Varid{t}} with
\ensuremath{\Conid{Sem}\;\Varid{m}\;\Varid{t}_{1}\to \Varid{m}\;(\Conid{Sem}\;\Varid{m}\;\Varid{t}_{2})} and each \ensuremath{\Conid{TInt}} with \ensuremath{\Conid{Int}}.

In order to make \ensuremath{\Conid{Sem}} into a proper type---as opposed to a mere type
synonym---and simultaneously add the monad \ensuremath{\Varid{m}} at the top level, we
define a \ensuremath{\mathbf{newtype}} \ensuremath{\Conid{M}}:
\begin{hscode}\SaveRestoreHook
\column{B}{@{}>{\hspre}l<{\hspost}@{}}%
\column{3}{@{}>{\hspre}l<{\hspost}@{}}%
\column{E}{@{}>{\hspre}l<{\hspost}@{}}%
\>[B]{}\mathbf{newtype}\;\Conid{M}\;\Varid{m}\;\Varid{i}\mathrel{=}\Conid{M}\;\{\mskip1.5mu \Varid{unM}\mathbin{::}\Varid{m}\;(\Conid{Sem}\;\Varid{m}\;\Varid{i})\mskip1.5mu\}{}\<[E]%
\\[\blanklineskip]%
\>[B]{}\mathbf{class}\;\Conid{Monad}\;\Varid{m}\Rightarrow \Conid{Eval}\;\Varid{m}\;\Varid{f}\;\mathbf{where}{}\<[E]%
\\
\>[B]{}\hsindent{3}{}\<[3]%
\>[3]{}\phi_{\mathrm{Eval}}\mathbin{::}\Varid{f}\;(\Conid{M}\;\Varid{m})\;(\Conid{M}\;\Varid{m})\;\Varid{i}\to \Varid{m}\;(\Conid{Sem}\;\Varid{m}\;\Varid{i})\mbox{\onelinecomment  \ensuremath{\Conid{M}\mathbin{.}\phi_{\mathrm{Eval}}\mathbin{::}\Conid{Alg}\;\Varid{f}\;(\Conid{M}\;\Varid{m})} is the actual algebra}{}\<[E]%
\\[\blanklineskip]%
\>[B]{}\Varid{eval}\mathbin{::}(\Conid{Monad}\;\Varid{m},\Conid{HDifunctor}\;\Varid{f},\Conid{Eval}\;\Varid{m}\;\Varid{f})\Rightarrow \Conid{Term}\;\Varid{f}\;\Varid{i}\to \Varid{m}\;(\Conid{Sem}\;\Varid{m}\;\Varid{i}){}\<[E]%
\\
\>[B]{}\Varid{eval}\mathrel{=}\Varid{unM}\mathbin{.}\Varid{cata}\;(\Conid{M}\mathbin{.}\phi_{\mathrm{Eval}}){}\<[E]%
\ColumnHook
\end{hscode}\resethooks

We can then provide the instance declarations for the signatures of
the core language, and effectively obtain a tagless, modular, and
extendable monadic interpreter:
\begin{hscode}\SaveRestoreHook
\column{B}{@{}>{\hspre}l<{\hspost}@{}}%
\column{3}{@{}>{\hspre}l<{\hspost}@{}}%
\column{37}{@{}>{\hspre}l<{\hspost}@{}}%
\column{38}{@{}>{\hspre}l<{\hspost}@{}}%
\column{E}{@{}>{\hspre}l<{\hspost}@{}}%
\>[B]{}\mathbf{instance}\;\Conid{Monad}\;\Varid{m}\Rightarrow \Conid{Eval}\;\Varid{m}\;\Conid{Lam}\;\mathbf{where}{}\<[E]%
\\
\>[B]{}\hsindent{3}{}\<[3]%
\>[3]{}\phi_{\mathrm{Eval}}\;(\Conid{Lam}\;\Varid{f})\mathrel{=}\Varid{return}\;(\Varid{unM}\mathbin{.}\Varid{f}\mathbin{.}\Conid{M}\mathbin{.}\Varid{return}){}\<[E]%
\\[\blanklineskip]%
\>[B]{}\mathbf{instance}\;\Conid{Monad}\;\Varid{m}\Rightarrow \Conid{Eval}\;\Varid{m}\;\Conid{App}\;\mathbf{where}{}\<[E]%
\\
\>[B]{}\hsindent{3}{}\<[3]%
\>[3]{}\phi_{\mathrm{Eval}}\;(\Conid{App}\;(\Conid{M}\;\Varid{mf})\;(\Conid{M}\;\Varid{mx}))\mathrel{=}\mathbf{do}\;{}\<[37]%
\>[37]{}\Varid{f}\leftarrow \Varid{mf}{}\<[E]%
\\
\>[37]{}\Varid{mx}\bind \Varid{f}{}\<[E]%
\\[\blanklineskip]%
\>[B]{}\mathbf{instance}\;\Conid{Monad}\;\Varid{m}\Rightarrow \Conid{Eval}\;\Varid{m}\;\Conid{Lit}\;\mathbf{where}{}\<[E]%
\\
\>[B]{}\hsindent{3}{}\<[3]%
\>[3]{}\phi_{\mathrm{Eval}}\;(\Conid{Lit}\;\Varid{n})\mathrel{=}\Varid{return}\;\Varid{n}{}\<[E]%
\\[\blanklineskip]%
\>[B]{}\mathbf{instance}\;\Conid{Monad}\;\Varid{m}\Rightarrow \Conid{Eval}\;\Varid{m}\;\Conid{Plus}\;\mathbf{where}{}\<[E]%
\\
\>[B]{}\hsindent{3}{}\<[3]%
\>[3]{}\phi_{\mathrm{Eval}}\;(\Conid{Plus}\;(\Conid{M}\;\Varid{mx})\;(\Conid{M}\;\Varid{my}))\mathrel{=}\mathbf{do}\;{}\<[38]%
\>[38]{}\Varid{x}\leftarrow \Varid{mx}{}\<[E]%
\\
\>[38]{}\Varid{y}\leftarrow \Varid{my}{}\<[E]%
\\
\>[38]{}\Varid{return}\;(\Varid{x}\mathbin{+}\Varid{y}){}\<[E]%
\\[\blanklineskip]%
\>[B]{}\mathbf{instance}\;\Conid{MonadError}\;\Conid{String}\;\Varid{m}\Rightarrow \Conid{Eval}\;\Varid{m}\;\Conid{Err}\;\mathbf{where}{}\<[E]%
\\
\>[B]{}\hsindent{3}{}\<[3]%
\>[3]{}\phi_{\mathrm{Eval}}\;\Conid{Err}\mathrel{=}\Varid{throwError}\;\text{\tt \char34 error\char34}{}\<[E]%
\ColumnHook
\end{hscode}\resethooks
With the above definition of \ensuremath{\Varid{eval}} we have, for instance, that \ensuremath{\Varid{eval}\;\Varid{e}\mathbin{::}\Conid{Either}\;\Conid{String}\;\Conid{Int}} evaluates to the value \ensuremath{\Conid{Right}\;\mathrm{4}}. Due to the
fact that we now have a typed language, the \ensuremath{\Conid{Err}} constructor is the
only source of an erroneous computation---the interpreter cannot get
stuck. Moreover, since the modular specification of the interpreter
only enforces the constraint \ensuremath{\Conid{MonadError}\;\Conid{String}\;\Varid{m}} for the signature
\ensuremath{\Conid{Err}}, the term \ensuremath{\Varid{e}} can in fact be interpreted in the identity monad,
rather than the \ensuremath{\Conid{Either}\;\Conid{String}} monad, as it does not contain
$\textbf{error}$. Consequently, we know statically that the evaluation
of \ensuremath{\Varid{e}} cannot fail!

Note that computations over generalised PCDTs are not limited to the
tagless approach that we have illustrated above. We could have easily
reformulated the semantic domain \ensuremath{\Conid{Sem}\;\Varid{m}} from
Section~\ref{sec:monadic-interpretation} as a GADT to use it as the
carrier of a many-sorted algebra. Other natural carriers for
many-sorted algebras are the type families of terms \ensuremath{\Conid{Term}\;\Varid{f}}, of
course.

Other concepts that we have introduced for vanilla PCDTs before can be
transferred straightforwardly to generalised PCDTs in the same
fashion. This includes contexts and term homomorphisms.

\section{Practical Considerations}
\label{sec:pract-cons}

The motivation for introducing CDTs was to make Swierstra's
\dalc~\cite{swierstra08jfp} readily useful in practice. Besides
extending \dalc with various aspects, such as monadic computations and
term homomorphisms, the CDTs library provides all the generic
functionality as well as automatic derivation of boilerplate
code. With (generalised) PCDTs we have followed that path. Our library
provides Template Haskell~\cite{sheard02haskell} code to automatically
derive instances of the required type classes, such as \ensuremath{\Conid{Difunctor}} and
\ensuremath{\Conid{Ditraversable}}, as well as smart constructors and lifting of algebra
type classes to coproducts. Moreover, our library supports automatic
derivation of standard type classes \ensuremath{\Conid{Show}}, \ensuremath{\Conid{Eq}}, and \ensuremath{\Conid{Ord}} for terms,
similar to Haskell's \ensuremath{\mathbf{deriving}} mechanism. We show how to derive
instances of \ensuremath{\Conid{Eq}} in the following subsection. \ensuremath{\Conid{Ord}} follows in the
same fashion, and \ensuremath{\Conid{Show}} follows an approach similar to the pretty
printer in Section~\ref{sec:parametric-algebras-and-terms}, but using
the monad \ensuremath{\Conid{FreshM}} that is also used to determine equality, as we
shall see below.

Figure~\ref{fig:pcdts-example} provides the complete
source code needed to implement our example language from
Section~\ref{sec:motivating-example}. Note that we have derived
\ensuremath{\Conid{Show}}, \ensuremath{\Conid{Eq}}, and \ensuremath{\Conid{Ord}} instances for terms of the language---in
particular the term \ensuremath{\Varid{e}} is printed as \texttt{Let (Lit 2) (\char`\\ a
  -> App (Lam (\char`\\ b -> Plus b a)) (Lit 3))}.

\begin{figure}[t]
\begin{lstlisting}[linerange={6-87},language=lithaskell,basicstyle=\tiny\ttfamily]
import Data.Comp.Param
import Data.Comp.Param.Show ()
import Data.Comp.Param.Equality ()
import Data.Comp.Param.Ordering ()
import Data.Comp.Param.Derive
import Control.Monad.Error (MonadError, throwError)

data Lam a b  = Lam (a -> b)
data App a b  = App b b
data Lit a b  = Lit Int
data Plus a b = Plus b b
data Let a b  = Let b (a -> b)
data Err a b  = Err

$(derive [smartConstructors, makeDifunctor, makeShowD, makeEqD, makeOrdD]
         [''Lam, ''App, ''Lit, ''Plus, ''Let, ''Err])

e :: Term (Lam :+: App :+: Lit :+: Plus :+: Let :+: Err)
e = Term (iLet (iLit 2) (\x -> (iLam (\y -> y `iPlus` x) `iApp` iLit 3)))

-- * Desugaring
class Desug f g where
  desugHom :: Hom f g

$(derive [liftSum] [''Desug]) -- lift Desug to coproducts

desug :: (Difunctor f, Difunctor g, Desug f g) => Term f -> Term g
desug (Term t) = Term (appHom desugHom t)

instance (Difunctor f, Difunctor g, f :<: g) => Desug f g where
  desugHom = In . fmap Hole . inj -- default instance for core signatures

instance (App :<: f, Lam :<: f) => Desug Let f where
  desugHom (Let e1 e2) = inject (Lam (Hole . e2)) `iApp` Hole e1

-- * Constant folding
class Constf f g where
  constfAlg :: forall a. Alg f (Trm g a)

$(derive [liftSum] [''Constf]) -- lift Constf to coproducts

constf :: (Difunctor f, Constf f g) => Term f -> Term g
constf t = Term (cata constfAlg t)

instance (Difunctor f, f :<: g) => Constf f g where
  constfAlg = inject . dimap Var id -- default instance

instance (Plus :<: f, Lit :<: f) => Constf Plus f where
  constfAlg (Plus e1 e2) = case (project e1, project e2) of
                             (Just (Lit n),Just (Lit m)) -> iLit (n + m)
                             _                           -> e1 `iPlus` e2

-- * Call-by-value evaluation
data Sem m = Fun (Sem m -> m (Sem m)) | Int Int

class Monad m => Eval m f where
  evalAlg :: Alg f (m (Sem m))

$(derive [liftSum] [''Eval]) -- lift Eval to coproducts

eval :: (Difunctor f, Eval m f) => Term f -> m (Sem m)
eval = cata evalAlg

instance Monad m => Eval m Lam where
  evalAlg (Lam f) = return (Fun (f . return))

instance MonadError String m => Eval m App where
  evalAlg (App mx my) = do x <- mx
                           case x of Fun f -> my >>= f
                                     _     -> throwError "stuck"

instance Monad m => Eval m Lit where
  evalAlg (Lit n) = return (Int n)

instance MonadError String m => Eval m Plus where
  evalAlg (Plus mx my) = do x <- mx
                            y <- my
                            case (x,y) of (Int n,Int m) -> return (Int (n + m))
                                          _             -> throwError "stuck"

instance MonadError String m => Eval m Err where
  evalAlg Err = throwError "error"
\end{lstlisting}
  \caption{Complete example using the parametric compositional data
    types library.}
  \label{fig:pcdts-example}
\end{figure}

\subsection{Equality}
\label{sec:pract-equality}

A common pattern when programming in Haskell is to derive instances of
the type class \ensuremath{\Conid{Eq}}, for instance in order to test the desugaring
transformation in Section~\ref{sec:term-transformations}. While the
use of PHOAS ensures that all functions are invariant under
$\alpha$-renaming, we still have to devise an algorithm that decides
$\alpha$-equivalence. To this end, we will turn the rather elusive
representation of bound variables via functions into a concrete form.

In order to obtain concrete representations of bound variables, we
provide a method for generating fresh variable names. This is achieved via
a monad \ensuremath{\Conid{FreshM}} offering the following operations:
\begin{hscode}\SaveRestoreHook
\column{B}{@{}>{\hspre}l<{\hspost}@{}}%
\column{13}{@{}>{\hspre}l<{\hspost}@{}}%
\column{E}{@{}>{\hspre}l<{\hspost}@{}}%
\>[B]{}\Varid{withName}\mathbin{::}(\Conid{Name}\to \Conid{FreshM}\;\Varid{a})\to \Conid{FreshM}\;\Varid{a}{}\<[E]%
\\[\blanklineskip]%
\>[B]{}\Varid{evalFreshM}{}\<[13]%
\>[13]{}\mathbin{::}\Conid{FreshM}\;\Varid{a}\to \Varid{a}{}\<[E]%
\ColumnHook
\end{hscode}\resethooks
\ensuremath{\Conid{FreshM}} is an abstraction of an infinite sequence of fresh
names. The function \ensuremath{\Varid{withName}} provides a fresh name. Names
are represented by the abstract type \ensuremath{\Conid{Name}}, which implements instances
of \ensuremath{\Conid{Show}}, \ensuremath{\Conid{Eq}}, and \ensuremath{\Conid{Ord}}.

We first introduce a variant of the type class \ensuremath{\Conid{Eq}} that uses the
\ensuremath{\Conid{FreshM}} monad:
\begin{hscode}\SaveRestoreHook
\column{B}{@{}>{\hspre}l<{\hspost}@{}}%
\column{3}{@{}>{\hspre}l<{\hspost}@{}}%
\column{E}{@{}>{\hspre}l<{\hspost}@{}}%
\>[B]{}\mathbf{class}\;\Conid{PEq}\;\Varid{a}\;\mathbf{where}{}\<[E]%
\\
\>[B]{}\hsindent{3}{}\<[3]%
\>[3]{}\Varid{peq}\mathbin{::}\Varid{a}\to \Varid{a}\to \Conid{FreshM}\;\Conid{Bool}{}\<[E]%
\ColumnHook
\end{hscode}\resethooks
This type class is used to define the type class \ensuremath{\Conid{EqD}} of equatable
difunctors, which lifts to coproducts:
\begin{hscode}\SaveRestoreHook
\column{B}{@{}>{\hspre}l<{\hspost}@{}}%
\column{3}{@{}>{\hspre}l<{\hspost}@{}}%
\column{16}{@{}>{\hspre}l<{\hspost}@{}}%
\column{25}{@{}>{\hspre}c<{\hspost}@{}}%
\column{25E}{@{}l@{}}%
\column{28}{@{}>{\hspre}l<{\hspost}@{}}%
\column{E}{@{}>{\hspre}l<{\hspost}@{}}%
\>[B]{}\mathbf{class}\;\Conid{EqD}\;\Varid{f}\;\mathbf{where}{}\<[E]%
\\
\>[B]{}\hsindent{3}{}\<[3]%
\>[3]{}\Varid{eqD}\mathbin{::}\Conid{PEq}\;\Varid{a}\Rightarrow \Varid{f}\;\Conid{Name}\;\Varid{a}\to \Varid{f}\;\Conid{Name}\;\Varid{a}\to \Conid{FreshM}\;\Conid{Bool}{}\<[E]%
\\[\blanklineskip]%
\>[B]{}\mathbf{instance}\;(\Conid{EqD}\;\Varid{f},\Conid{EqD}\;\Varid{g})\Rightarrow \Conid{EqD}\;(\Varid{f}\fplus\Varid{g})\;\mathbf{where}{}\<[E]%
\\
\>[B]{}\hsindent{3}{}\<[3]%
\>[3]{}\Varid{eqD}\;(\Conid{Inl}\;\Varid{x})\;{}\<[16]%
\>[16]{}(\Conid{Inl}\;\Varid{y}){}\<[25]%
\>[25]{}\mathrel{=}{}\<[25E]%
\>[28]{}\Varid{x}\mathbin{`\Varid{eqD}`}\Varid{y}{}\<[E]%
\\
\>[B]{}\hsindent{3}{}\<[3]%
\>[3]{}\Varid{eqD}\;(\Conid{Inr}\;\Varid{x})\;{}\<[16]%
\>[16]{}(\Conid{Inr}\;\Varid{y}){}\<[25]%
\>[25]{}\mathrel{=}{}\<[25E]%
\>[28]{}\Varid{x}\mathbin{`\Varid{eqD}`}\Varid{y}{}\<[E]%
\\
\>[B]{}\hsindent{3}{}\<[3]%
\>[3]{}\Varid{eqD}\;\anonymous \;{}\<[16]%
\>[16]{}\anonymous {}\<[25]%
\>[25]{}\mathrel{=}{}\<[25E]%
\>[28]{}\Varid{return}\;\Conid{False}{}\<[E]%
\ColumnHook
\end{hscode}\resethooks
We then obtain equality of terms as follows (we do not consider
contexts here for simplicity):
\begin{hscode}\SaveRestoreHook
\column{B}{@{}>{\hspre}l<{\hspost}@{}}%
\column{5}{@{}>{\hspre}l<{\hspost}@{}}%
\column{20}{@{}>{\hspre}l<{\hspost}@{}}%
\column{31}{@{}>{\hspre}l<{\hspost}@{}}%
\column{E}{@{}>{\hspre}l<{\hspost}@{}}%
\>[B]{}\mathbf{instance}\;\Conid{EqD}\;\Varid{f}\Rightarrow \Conid{PEq}\;(\Conid{Trm}\;\Varid{f}\;\Conid{Name})\;\mathbf{where}{}\<[E]%
\\
\>[B]{}\hsindent{5}{}\<[5]%
\>[5]{}\Varid{peq}\;(\Conid{In}\;\Varid{t}_{\mathrm{1}})\;{}\<[20]%
\>[20]{}(\Conid{In}\;\Varid{t}_{\mathrm{2}}){}\<[31]%
\>[31]{}\mathrel{=}\Varid{t}_{\mathrm{1}}\mathbin{`\Varid{eqD}`}\Varid{t}_{\mathrm{2}}{}\<[E]%
\\
\>[B]{}\hsindent{5}{}\<[5]%
\>[5]{}\Varid{peq}\;(\Conid{Var}\;\Varid{x}_{\mathrm{1}})\;{}\<[20]%
\>[20]{}(\Conid{Var}\;\Varid{x}_{\mathrm{2}}){}\<[31]%
\>[31]{}\mathrel{=}\Varid{return}\;(\Varid{x}_{\mathrm{1}}\equiv \Varid{x}_{\mathrm{2}}){}\<[E]%
\\
\>[B]{}\hsindent{5}{}\<[5]%
\>[5]{}\Varid{peq}\;\anonymous \;{}\<[20]%
\>[20]{}\anonymous {}\<[31]%
\>[31]{}\mathrel{=}\Varid{return}\;\Conid{False}{}\<[E]%
\\[\blanklineskip]%
\>[B]{}\mathbf{instance}\;(\Conid{Difunctor}\;\Varid{f},\Conid{EqD}\;\Varid{f})\Rightarrow \Conid{Eq}\;(\Conid{Term}\;\Varid{f})\;\mathbf{where}{}\<[E]%
\\
\>[B]{}\hsindent{5}{}\<[5]%
\>[5]{}(\equiv )\;(\Conid{Term}\;\Varid{x})\;(\Conid{Term}\;\Varid{y})\mathrel{=}\Varid{evalFreshM}\;((\Varid{x}\mathbin{::}\Conid{Trm}\;\Varid{f}\;\Conid{Name})\mathbin{`\Varid{peq}`}\Varid{y}){}\<[E]%
\ColumnHook
\end{hscode}\resethooks
Note that we need to explicitly instantiate the parametric type in \ensuremath{\Varid{x}}
to \ensuremath{\Conid{Name}} in the last instance declaration, in order to trigger the
instance for \ensuremath{\Conid{Trm}\;\Varid{f}\;\Conid{Name}} defined above.

Equality of terms, i.e.\ $\alpha$-equivalence, has thus been reduced
to providing instances of \ensuremath{\Conid{EqD}} for the difunctors comprising the
signature of the term, which for \ensuremath{\Conid{Lam}} can be defined as follows:
\begin{hscode}\SaveRestoreHook
\column{B}{@{}>{\hspre}l<{\hspost}@{}}%
\column{3}{@{}>{\hspre}l<{\hspost}@{}}%
\column{E}{@{}>{\hspre}l<{\hspost}@{}}%
\>[B]{}\mathbf{instance}\;\Conid{EqD}\;\Conid{Lam}\;\mathbf{where}{}\<[E]%
\\
\>[B]{}\hsindent{3}{}\<[3]%
\>[3]{}\Varid{eqD}\;(\Conid{Lam}\;\Varid{f})\;(\Conid{Lam}\;\Varid{g})\mathrel{=}\Varid{withName}\;(\lambda \Varid{x}\to \Varid{f}\;\Varid{x}\mathbin{`\Varid{peq}`}\Varid{g}\;\Varid{x}){}\<[E]%
\ColumnHook
\end{hscode}\resethooks
That is, \ensuremath{\Varid{f}} and \ensuremath{\Varid{g}} are considered equal if they are equal when
applied to the same fresh name \ensuremath{\Varid{x}}.

\section{Discussion and Related Work}
\label{sec:discussion}

Implementing languages with binders can be a difficult task. Using
explicit variable names, we have to be careful in order to make sure
that functions on ASTs are invariant under $\alpha$-renaming.
HOAS~\cite{pfenning88pldi} is one way of tackling this problem, by
reusing the binding mechanisms of the implementation language to
define those of the object language. The challenge with HOAS, however,
is that it is difficult to perform recursive computations over ASTs
with
binders~\cite{fegaras96popl,meijer95fpca,schuermann01tcs,washburn08jfp}.
Besides what is documented in this paper, we have also lifted
(generalised) parametric compositional data types to other
(co)recursion schemes, such as anamorphisms and
histomorphisms. Moreover, term homomorphisms can be straightforwardly
extended with a state space: depending on how the state is propagated,
this yields bottom-up resp.\ top-down tree
transducers~\cite{comon07book}.

Our approach of using PHOAS~\cite{chlipala08icfp} amounts to the same
restriction on embedded functions as Fegeras and
Sheard~\cite{fegaras96popl}, and Washburn and
Weirich~\cite{washburn08jfp}. However, unlike Washburn and Weirich's
Haskell implementation, our approach does not rely on making the type
of terms abstract. Not only is it interesting to see that we can do
without type abstraction, in fact, we sometimes need to inspect terms
in order to write functions that produce terms, such as our constant
folding algorithm. With Washburn and Weirich's encoding this is not
possible.

Ahn and Sheard~\cite{ahn11icfp} recently showed how to generalise the
recursion schemes of Washburn and Weirich to Mendler-style recursion
schemes, using the same representation for terms as Washburn and
Weirich. Hence their approach also suffers from the inability to
inspect terms. Although we could easily adopt Mendler-style recursion
schemes in our setting, their generality does not make a difference in
a non-strict language such as Haskell. Additionally, Ahn and Sheard
pose the open question whether there is a safe (i.e., terminating) way
to apply histomorphisms to terms with negative recursive occurrences:
although we have not investigated termination properties of our
histomorphisms, we conjecture that the use of our parametric
terms---which are purely inductive---may provide one solution.

The \emph{finally tagless} approach of Carette et
al.~\cite{carette09jfp} has been proposed as an alternative solution
to the expression problem~\cite{wadler98exp}. While the approach is
very simple and elegant, and also supports (typed) higher-order
encodings, the approach falls short when we want to define recursive,
modular computations that construct modular terms too. Atkey et
al.~\cite{atkey09haskell}, for instance, use the finally tagless
approach to build a modular interpreter. However, the interpreter
cannot be made modular in the return type, i.e.\ the language defining
values. Hence, when Atkey et al.\ extend their expression language
they need to also change the data type that represents values, which
means that the approach is not fully modular. Although our
interpreter in Section~\ref{sec:monadic-interpretation} also uses a
fixed domain of values \ensuremath{\Conid{Sem}}, we can make the interpreter fully
modular by also using a PCDT for the return type, and using a
multi-parameter type class definition similar to the desugaring
transformation in Section~\ref{sec:term-transformations}.

\emph{Nominal sets}~\cite{pitts06jacm} is another approach for dealing
with binders, in which variables are explicit, but recursively defined
functions are guaranteed to be invariant with respect to
$\alpha$-equivalence of terms. Implementations of this approach,
however, require extensions of the metalanguage~\cite{shinwell03icfp},
and the approach is therefore not immediately usable in Haskell.

\section*{Acknowledgement}
\label{sec:acknowledgement}

The authors wish to thank Andrzej Filinski for his insightful comments
on an earlier version of this paper.

\bibliographystyle{eptcs}
\bibliography{refs}

\end{document}